\documentclass{emulateapj}

\newcommand{\kms}{{\mathrm{km~s^{-1}}}}
\newcommand{\ergs}{{\mathrm{erg~s^{-1}}}}
\newcommand{\ergscm}{{\mathrm{erg~s^{-1}~cm^{-2}}}}

\newcommand{\lumcgs}{{\mathrm{erg~s^{-1}}}}

\newcommand{\dd}{\mathrm{d}}
\newcommand{\msun}{\mathrm{M}_\odot}
\newcommand{\sfr}{\mathrm{M_\odot~yr^{-1}}}
\newcommand{\mpc}{\mathrm{Mpc}}

\newcommand{\lya}{Lyman~$\alpha$}
\newcommand{\amsq}{{\mathrm{arcmin^{2}}}}
\newcommand{\llim}{L_\mathrm{lim}(\mathbf{\Omega},z)}
\newcommand{\mg}{\mathcal{M}(\mathbf{\Omega},z)}
\newcommand{\flim}{f_\mathrm{lim}(z)}
\newcommand{\slit}{T(\mathbf{\Omega})}
\newcommand{\sky}{\mathbf{\Omega}}

\slugcomment{\date}

\shorttitle{Low-luminosity Lyman $\alpha$ sources at high redshift}
\shortauthors{M. R. Santos et al.}

\begin{document}

\title{The Abundance of Low-luminosity Lyman $\alpha$ Emitters at High
Redshift\altaffilmark{1}}

\author{Michael R. Santos\altaffilmark{2}, Richard S. Ellis, Jean-Paul
Kneib\altaffilmark{3}} \affil{California Institute of Technology}
\affil{105-24 Caltech, Pasadena, CA 91125}
\email{mrs@tapir.caltech.edu}

\author{Johan Richard}
\affil{Observatoire de Midi-Pyr\'en\'ees}
\affil{UMR5572, 14 Av. Edouard B\'elin, F-31400 Toulouse, France}

\and

\author{Konrad Kuijken}
\affil{Sterrewacht Leiden}
\affil{P.O. Box 9513, NL-2300 RA, Leiden, The Netherlands}

\altaffiltext{1}{Data presented herein were obtained at the W.M. Keck
Observatory, which is operated as a scientific partnership among the
California Institute of Technology, the University of California and
the National Aeronautics and Space Administration. The Observatory was
made possible by the generous financial support of the W.M. Keck
Foundation.}

\altaffiltext{2}{current address Institute of Astronomy, University of Cambridge, Madingley Road, Cambridge CB1 1BL, U.K.}

\altaffiltext{3}{also Observatoire de Midi-Pyr\'en\'ees, UMR5572, 14
Av. Edouard B\'elin, F-31400 Toulouse, France}

\begin{abstract}
We derive the luminosity function of high-redshift \lya\ emitting
sources from a deep, blind, spectroscopic survey that utilized
strong-lensing magnification by intermediate-redshift clusters of
galaxies.  We observed carefully selected regions near 9 clusters,
consistent with magnification factors generally greater than 10 for
the redshift range 4.5$<z<$6.7.  Eleven emission-line candidates were
located in the range 2.2$<z<$5.6 whose identification we justify as
\lya, in most cases via further spectroscopic observations.  The
selection function we constructed for our survey takes into account
our varying intrinsic \lya\ line sensitivity as a function of
wavelength and sky position.  By virtue of the strong magnification
factor, we provide constraints on the \lya\ luminosity function to
unprecedented limits of $10^{40}~\ergs$, corresponding to a
star-formation rate of $0.01~\sfr$.  Our cumulative $z\simeq5$ \lya\
luminosity function is consistent with a power law form, $n(>L)\propto
L^{-1}$ over $10^{41}$ to $10^{42.5}~\ergs$.  When combined with the
results of other surveys, limited at higher luminosities, our results
suggest evidence for the suppression of star formation in low-mass
halos, as predicted in popular models of galaxy formation.
\end{abstract}


\section{Introduction}

The epoch of cosmic reionization, when the intergalactic hydrogen in
the universe transitioned from neutral to ionized, is the current
frontier of observational cosmology.  QSOs discovered by the Sloan
Digital Sky Survey (SDSS) indicate that reionization was just
finishing at $z \simeq 6$ \citep{bec01,djo01,fan02}.  Recent results
from the \textit{WMAP} satellite suggest that significant reionization
of the universe took place by $z\sim12$ \citep{spe03}.  The sources
that reionized the universe, however, are still unknown: at $z\sim6$
neither bright QSOs discovered by SDSS \citep{fan01a} nor faint AGN
from deep x-ray observations \citep{barg03} produced enough photons to
reionize the universe.  Other evidence from the temperature and
ionization state of the intergalactic medium (IGM) suggests that,
though QSOs dominated the meta-galactic ionizing background at $z \sim
3$, the spectrum was softer at reionization \citep[e.g.][]{sok03}.
Accordingly, hot stars in early star-forming systems may have been the
dominant source of reionizing photons.  One goal of the forthcoming
NASA/ESA \textit{James Webb Space Telescope} (\textit{JWST}), a
6-meter IR telescope scheduled for launch in 2010, is to study the
formation of the first generations of galaxies and their contribution
to reionization \citep{mat00}.

Early galaxies played many important roles beyond their involvement
with reionization.  The IGM was enriched well above the primordial
metal abundance by $z=5$ \citep{son01,pet03}; additional evidence for
early metal production comes from metal-poor globular clusters in the
Milky Way.  Age estimates imply a formation epoch of $z \ga 4$ for
current cosmological parameters \citep{kra03}, but the typical
metallicity of these objects is $10^{-2}$ of the solar value
\citep{har96}.  The stars responsible for reionization and early metal
production may still be present in some form today.  It is an
important challenge to identify the transition between the very first,
metal-free, stars, and \mbox{Population II} stars, because of the
strong constraints on the metallicity of low-mass stars provided by
studies of halo stars in the Milky Way.  A complete understanding of
the metallicity distribution of old Galactic stars will benefit from
direct observation of very high redshift star formation in
\textit{proto-galactic systems that will evolve into galaxies like the
Milky Way}.

In advance of \textit{JWST}, which will use IR capabilities to observe
galaxies before the end of reionization in rest-frame UV and optical
light, current ground-based facilities have the opportunity to
discover and characterize star-forming galaxies near the epoch of
reionization with rest-frame UV observations.  In particular, the
identification of \lya\ emission from star-forming regions of early
galaxies has proven to be a powerful tool for discovering $z>4$
galaxies and measuring their redshifts (see Section~\ref{sec:lyasur}).
The redshift range $5<z<7$ is of particular interest for two reasons.
One is that the very detection of \lya\ emission may place a
constraint on the progress of reionization
\citep{rho01,hai02,hu02a,rho03}: it is difficult to observe \lya\
emission from galaxies embedded in a neutral IGM, but the strength of
the constraint derived from the successful detection of \lya\ depends
on the assumed properties of the sources \citep{san03}.  The second
important reason to study galaxy formation during and after
reionization is that an intense UV background and $10^4$~K IGM is
predicted to suppress star-formation in galaxies that form after
reionization \citep{bar99,gne00,ben02a}.  There is a discrepancy
between some theoretical predictions of the abundance of dark matter
halos on dwarf-galaxy mass scales and the number of observed dwarf
satellites in the Local Group \citep{moo99}.  Reionization may
sterilize many dwarf galaxy-scale halos to star formation, so that the
luminous satellites of the Milky Way are dwarf galaxies formed before
reionization, and the remaining satellite halos are empty of stars and
thus dark \citep{ben02b,som02}.

This paper presents the results of a spectroscopic \lya\ emission-line
survey for galaxies at $2.2<z<6.7$ that utilizes the strong-lensing
properties of intermediate-redshift clusters to magnify the surveyed
regions.  In Section~\ref{sec:lyasur} we review the use of \lya\
surveys as probes of early star formation.  Section~\ref{sec:lowlum}
motivates the importance of surveys for low-luminosity galaxies.
Section~\ref{sec:obs} describes the advantages of a survey utilizing
strong lensing and details our survey strategy, targets, observations,
and data reduction.  In Section~\ref{sec:detect} our \lya\
emission-line detections are presented.  We compute our survey volume
and source number density in
Section~\ref{sec:suranal}. Section~\ref{sec:comp} compares the results
of our survey to other surveys and theoretical models.  In
Section~\ref{sec:sum} we summarize.

Throughout this paper we use a $\Lambda$CDM
cosmological model with ($\Omega_\mathrm{m}, \Omega_\mathrm{\Lambda},
\Omega_\mathrm{b}, \sigma_8$)=($0.3,0.7,0.043,0.9$) and $h\equiv
H_0/(100~\kms\mpc^{-1})=0.7$; these values are consistent with the
values derived in \citet{spe03}.

\section{Lyman~$\alpha$ Surveys }
\label{sec:lyasur}

The primary appeal of \lya\ emission as a signpost to high-redshift
galaxy formation is that it traces star formation at a wavelength that
conveniently redshifts into the visible and near-IR, where sensitive,
high-angular resolution observations are currently most practical.
The \lya\ emission line may be quite strong, but its luminosity is
quite sensitive to physical details of the nature and geometry of the
star-forming regions.  Because \lya\ emission traces hydrogen
recombinations, it is intimately related to the production of ionizing
photons by the stars present.  Both the initial mass function (IMF)
and metallicity of the stars affect their production rate of ionizing
photons.  However, if the IMF and metallicity are constrained to
reasonable ranges, the ionizing photon production rate can be reliably
connected to the star-formation rate.

The major complication for the interpretation of \lya\ line strengths
is the effect of the nebula around the star-forming region.  Hydrogen
at low density does not recombine quickly, so, e.g., ionizing photons
that escape into the IGM are ``lost'' for the purposes of producing a
\lya\ emission line.  Even after a hydrogen recombination produces a
\lya\ photon, which happens for about two thirds of the recombinations
\citep{ost89}, there are many ways in which the photon may be
destroyed prior to escape.  The resonant nature of the \lya\
transition results in a very short mean free path, even in a mostly
ionized nebula.  Consequently, if dust is mixed in with the gas, then
the chance of absorption by a dust grain may be higher for a \lya\
photon than a non-resonantly scattered photon at the same wavelength
\citep[but see][for an alternative]{neu91}.  The dust content of very
high-redshift galaxies is still relatively unconstrained, and will
likely remain so at least until \textit{JWST}.

On the positive side, \lya\ is the intrinsically strongest
recombination line from an H II region.  Another meritorious aspect of
\lya\ is that its emission strength does not strongly depend on the
metallicity of the ionized gas (the only effect is from the
temperature of the photoionized gas, which depends on metallicity);
consequently, it can be used as a tracer for truly primordial star
formation, where dust extinction is also believed not to be a problem.

\citet{par67} introduced a model of galaxy formation ``to assess the
general possibility of observing distant, newly formed, galaxies.''
In their model they estimated that 6--7\% of the luminosity of early
galaxies would be emitted in the \lya\ emission line: they predicted
line luminosities of $2\times10^{45}~\mathrm{erg~s^{-1}}$ over galaxy
formation time-scales of $3\times10^{7}$~yr.  The predictions of
\citet{par67} and others led to many observational surveys for \lya\
emission from high-redshift galaxies, reviewed by \citet[\S
4.5]{pri94}.  \citeauthor{pri94} summarized the status of searches at
that time with ``no emission line primeval galaxies have been found''
despite 16 cited surveys covering various redshifts ranges from $z=2$
to $z=5$.  In striking contrast to these pioneering explorations, many
high-redshift \lya\ emission-line galaxies have been discovered and
confirmed in the past 8 years.  \citet[\S 4]{ste99} and \citet{tan03}
have reviewed recent progress, and we provide a brief summary here.

The search technique that has dominated recent success in the
discovery of large numbers of \lya\ emission-line galaxies at $z>4$ is
narrow-band photometry.  This approach uses a narrow ($\sim100$~\AA)
filter chosen to lie in a spectral region of low sky background; such
surveys cover relatively large areas of sky with sensitivity to \lya\
emission over a small window in redshift, $\Delta z \sim 0.1$.  Many
groups have now performed successful blind narrow-band surveys for
$z>4$ galaxies: \citet*{hu98} and \citet{rho00} at $z=4.5$;
\citet{ouc03} at $z=4.9$; \citet*{hu99}, \citet{rho01}, and
\citet{rho03} at $z=5.7$; and \citet{hu02a} and \citet{kod03} at
$z=6.5$.

Spectroscopic surveys provide a complementary technique to the
narrow-band method.  For equal observing time at one position,
spectroscopic searches at optical wavelengths can cover a large range
in redshift, $\Delta z \sim 3$, to better line flux sensitivity than a
corresponding narrow-band survey.  However, the area surveyed by a
long slit is typically $\sim 5\times10^{-2}~\mathrm{arcmin}^2$, in
contrast to 20--2000~$\mathrm{arcmin}^2$ for an imaging camera.  Deep
long slit observations of other targets have been searched for
serendipitous detection of high-redshift galaxies.  These techniques
discovered the first confirmed $z>5$ galaxy \citep{dey98} and
subsequently turned up a few more $z>4$ sources \citep{hu98,wey98b}.
Serendipitous surveys will likely continue to play a role in
discovering high-redshift galaxies, since the deepest spectra on the
largest telescopes are likely to be pointed observations rather than
devoted emission-line surveys.

Under sky-limited conditions, it is simple to show that the
signal-to-noise ratio $\cal R$ that is reached on pure emission-line
sources of flux $F$ in a survey over sky area $A$ and wavelength range
$\Lambda$, with area and wavelength coverage $\Delta A$ and
$\Delta\lambda$, seeing disk $\delta A$ and spectral resolution
$\delta\lambda$, is given by
\begin{equation}
{\cal R}={F t\over \sqrt{S\,\delta A\,\delta\lambda\, t}} =
   {F\sqrt{T/(A\Lambda)} \over \sqrt{ S {\delta A\over\Delta A}
   {\delta\lambda\over\Delta\Lambda}}},
\end{equation}
where $t$ and $T$ are the lengths of individual exposures, and of the
whole survey, respectively, and $S$ is the sky surface brightness per
unit wavelength. The tradeoff is thus between the number of seeing
elements that can be observed simultaneously ($O(N^2)$ in case of an
imaging survey, $O(N)$ for a long-slit spectral survey) vs.\ the
number of spectral elements (1 for a narrow-band survey, $O(N)$ for a
spectral survey). As each is limited by detector sizes, both modes
can in principle offer comparable survey speed to a given flux limit
(the argument applies also to surveys with integral field unit
spectrographs). Spectral surveys are advantageous for targeting small
regions of the sky and covering a large wavelength range, or for
simply going as deep as possible in a given amount of time;
narrow-band surveys are optimal in the opposite regime of wide area,
small wavelength range. A further advantage of high-spectral
resolution surveys is that the night sky emission lines can be
maximally avoided, which is not possible with the typical bandwidths
employed in narrow-band surveys.

A few other hybrid techniques combine aspects of the narrow-band
imaging and long-slit spectroscopy approaches.  \citet{mai03} used an
imaging Fabry-Perot interferometer to take sequences of narrow band
images within the night-sky windows corresponding to $z=4.8,5.7,6.5$,
and have confirmed discoveries at $z=4.8$ and 5.7.  Recently
\citet{lil03} and \citet{mar03} used a ``slit-let'' slit mask with a
narrow-band filter to do spectroscopic surveys over a relatively large
areas and narrow redshift windows.  There is also an ongoing
\textit{HST} program to use slitless spectroscopy with the ACS camera
grism to discover high-redshift \lya\ emission (J. E. Rhoads,
priv. comm.).

\section{Searches for Distant Low-luminosity Galaxies}
\label{sec:lowlum}

In this section we present the motivation for conducting a survey
devoted specifically to low-luminosity $z\sim5$ \lya\ emitting
galaxies ($\la10^{42}~\ergs$ in the \lya\ line), an unexplored region
of survey parameter space.

At $z \sim 5$, the luminous \lya\ galaxies (\S\ref{sec:lyasur}) and
QSOs \citep[e.g.,][]{fan99,fan00,fan01a} discovered so far represent
the rarest and most spectacular tail of the range of structure
formation scales (e.g., \citealt{bar03a} suggested that the high-$z$
SDSS QSOs reside in $10^{12}~\msun$ virialized halos).  They almost
certainly evolve into the rarest and most massive environments in the
local universe: rich clusters of galaxies.  In striking contrast, the
characteristic mass of virialized halos at $z\sim5$ is only
$10^9$\,M$_\odot$; if such an object steadily converts its
$\sim10^8$\,M$_\odot$ of baryons into stars, it will have a
star-formation rate of only $\sim0.1\,\mathrm{M_\odot\,yr^{-1}}$.

These objects, which we will refer to as low-mass halos, would not be
detectable in any of the surveys cited above, but our understanding of
galaxy formation depends crucially on constraining their properties
for three important reasons: First, they represent the most common
environment \emph{by mass} of virialized halos (the peak of the
mass-weighted mass function is always near the characteristic mass);
consequently, if low-mass halos form stars efficiently, they could
dominate the star-formation rate at high redshift.  Second, they are
the progenitors of common galaxies in poor environments, like the
Milky Way, under the current paradigm of structure formation.  The
detection of low-mass sources is a direct test of the ``bottom-up''
description of galaxy assembly.  Third, low-mass objects have a unique
link to the IGM: kinetic energy injected into the IGM by
photoionization at $z\ga6$ is expected to raise the cosmic Jeans mass
and inhibit gas cooling; this effect has no consequence for the
brightest sources residing in deep potential wells, but may heavily
suppress star formation in $10^9$\,M$_\odot$ objects
\citep[e.g.,][]{bar99,gne00}.  This Jeans-mass effect has been cited
as the solution for the Cold Dark Matter ``crisis'' of over-predicting
the number of Milky Way satellite galaxies compared to observation
\citep{ben02b,som02}.  Low-mass halos also place relevant constraints
on the energy scale of dark matter in Warm Dark Matter models
(Z. Haiman, priv. comm.).

A practical and strategic advantage in characterizing the luminosity
function at low luminosities is to determine the optimum survey depth
for future surveys that aim to discover large numbers of $4\la z\la7$
galaxies.  If the luminosity function were very steep, then deep
surveys such as ours would be more efficient than shallower, wider
field surveys.  Theoretical prejudice suggests that the luminosity
function should have a steep effective slope in the region associated
with the exponential cut-off of the number density of underlying
halos, assuming some sort of mass-to-luminosity correspondence.
Previous non-detections combined with recent successes seem to bear
this out \citep[and see below]{pri94}.  That is, current surveys may
be approaching the characteristic luminosity; however, the
characteristic luminosity and luminosity function shape have yet to be
well constrained.  In the local universe, luminosity functions based
on star-formation rate estimators such as H$\alpha$ luminosity roughly
follow the Schechter function \citep{sch76} form of power-law behavior
at low luminosity, reflecting the underlying power-law of the mass
function, albeit with possibly a different slope
\citep[e.g.,][]{gal95}.

The predicted suppression of star formation in halos with small
potential wells suggests that the \lya\ luminosity function at $z \sim
5$ may have a modified shape.  The reason is that the characteristic
mass scale where the exponential tail and power law regions of the
mass scale meet, $\sim 10^9~\msun$, corresponds quite closely to the
mass scale where a number of physical mechanisms may suppress star
formation.  We described the effect of a hot IGM above.  In addition,
energy injected in the ISM of star-forming galaxies by stellar winds
and supernovae (called ``negative feedback,'' or just ``feedback'') is
predicted to heavily suppress star formation in halos with circular
velocities below about $100~\kms$ \citep{dek86}, corresponding to a
mass scale at $z \sim 5$ of $\sim 10^{11}~\msun$.  A current
implementation of feedback by \citet{ben02a} shows the importance of
feedback on the high-redshift star-formation rate.  This effect
complements the inhibiting effects of a hot IGM on star formation in
low-mass halos.  A third mechanism that may reduce star formation
preferentially in low-mass halos is the effect of large-scale winds
blown by star-forming galaxies.  \citet{sca00} and \citet{sca01}
computed the influence of winds blown out through the IGM by the first
galaxies to form.  They concluded that these winds effectively sweep
gas out of nearby halos in the process of collapsing, meaning that
even though the dark matter continues its collapse to virialization,
there is little corresponding star formation because of the lack of
baryons.  In their model winds influence the entire star-formation
history of the universe, but at $z \sim 5$ particularly suppress
star-formation in halos smaller than $\sim 10^{10}~\msun$.

Figure~\ref{fig:schemec} is a schematic illustration of the possible
effect of the suppression of star formation in low-mass halos.  The
solid curve is the mass function of halos at $z=5$ \citep{shet02}, and
we have converted the mass scale on the top axis into a star-formation
rate on the bottom axis using a simple prescription (see
Section~\ref{sec:halo}).  In this simple model, the \lya\ line
luminosity function, if interpreted as directly proportional to that
of the SFR, would have the shape of the solid curve.  We have
introduced the suppression of star formation in low mass halos by
using the prescription of \citet{gne00} to efficiently filter out gas
from halos below a critical mass scale $M_\mathrm{F}$, the filtering
mass \citep{gne00}.  Each of the broken curves is the shape of the
\lya\ line luminosity function we expect (again assuming it scales
with SFR) after applying filtering on a different mass scale.
Ultimately the filtering mass, and thus the physics described in the
previous paragraph, may be constrained directly by a measurement of
the shape of the star-formation rate function, along with estimates of
the corresponding halo masses (see Section~\ref{sec:halo}).

Only recently have deep observations with large telescopes even
detected $z\sim5$ \lya\ emitting galaxies; the detailed form of the
luminosity function will not be well-constrained in the immediate
future.  However, our low-luminosity \lya\ survey, and others like it,
in concert with surveys at higher luminosities, may constrain or
detect the break in the luminosity function associated with the
characteristic halo mass and where star formation is suppressed.

\section{Observations}
\label{sec:obs}

\subsection{Survey Stategy}

The goal of our survey is to extend the census of \lya\ sources at $z
\sim 5$ to the faintest luminosities possible with existing
observational facilities.  We achieved this through deep spectroscopic
exposures on areas of sky strongly magnified by gravitational lensing.

Strong gravitational lensing by rich clusters of galaxies at $z \sim
0.2$ is an invaluable resource to a survey covering a very small area
of sky to great depth, as such clusters magnify background
high-redshift sources by greater than a factor of 10 over regions of
roughly $0.1~\amsq$ in the image plane.  This advantage comes at a
price: lensing increases the apparent area of a background source at
fixed surface brightness, so that the pointing is deeper by a factor
of the magnification, $\mathcal{M}$, but covers only $1/\mathcal{M}$
of the area of an equivalent unlensed pointing (assuming unresolved
sources). This is a superior strategy for a deep, narrow survey.
Achieving the same depth and area in an unlensed survey would require
a factor of $\mathcal{M}$ more observing time, a huge difference for
$\mathcal{M}\sim10$.

Only a small area of the whole sky, $\sim 100~\amsq$, is magnified by
a factor of 10 or more by clusters; this sets a fundamental limit to
the areal coverage of any survey utilizing strong lensing by clusters.
For a large survey that detected many sources, statistical information
about the distribution of magnification over the survey area might be
sufficient to construct a \lya\ emitter luminosity function.  However,
since we planned to survey a small area and detect only a few sources,
we chose to observe clusters with \textit{HST} imaging and
spectroscopic redshifts for many of the identified arcs and multiple
images \citep[e.g.,][]{kne96}.  These are currently available for only
a small fraction of strong-lensing clusters.  The positions and
spectroscopic redshifts of the arcs constrain the distribution of mass
in the cluster which can, in turn, be used to predict a magnification
map of the cluster for high-redshift sources.  Since lensing depends
on the angular-diameter distance between the source and the lens, and
that distance changes slowly with redshift at $z \sim 5$, the sky area
with large magnification is fairly independent of the source redshift
for $z>3$

\begin{figure}
  \includegraphics[width=7cm]{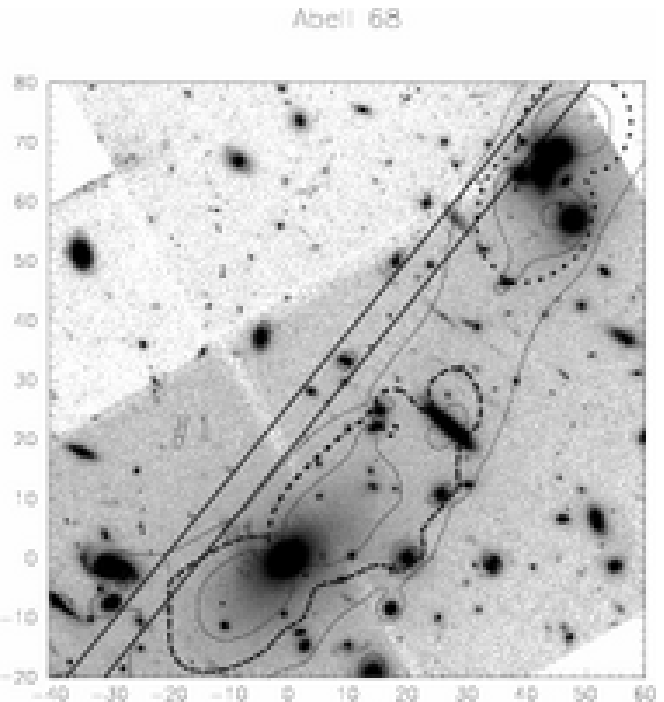}\includegraphics[width=7cm]{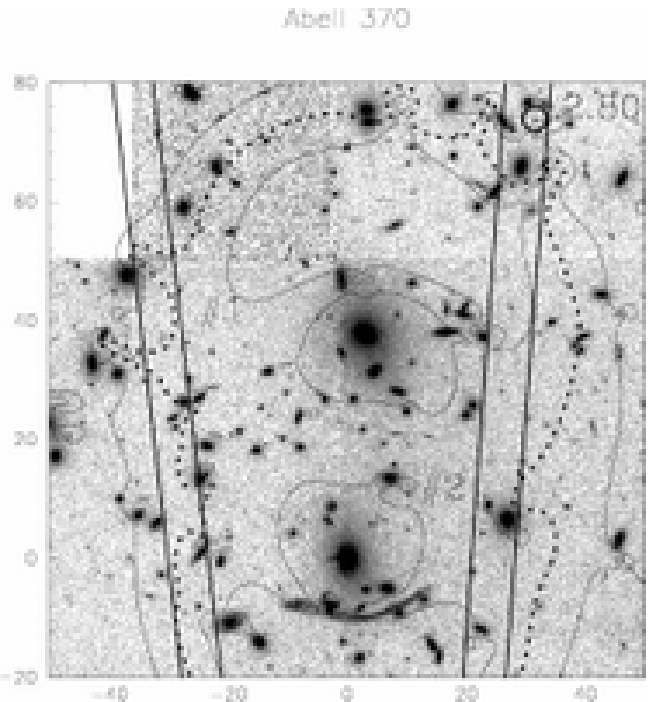}
  \includegraphics[width=7cm]{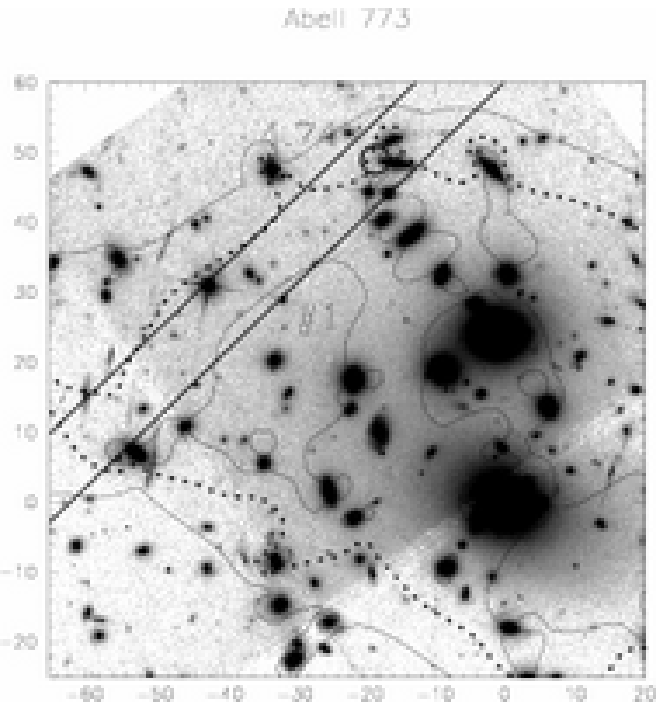}\includegraphics[width=7cm]{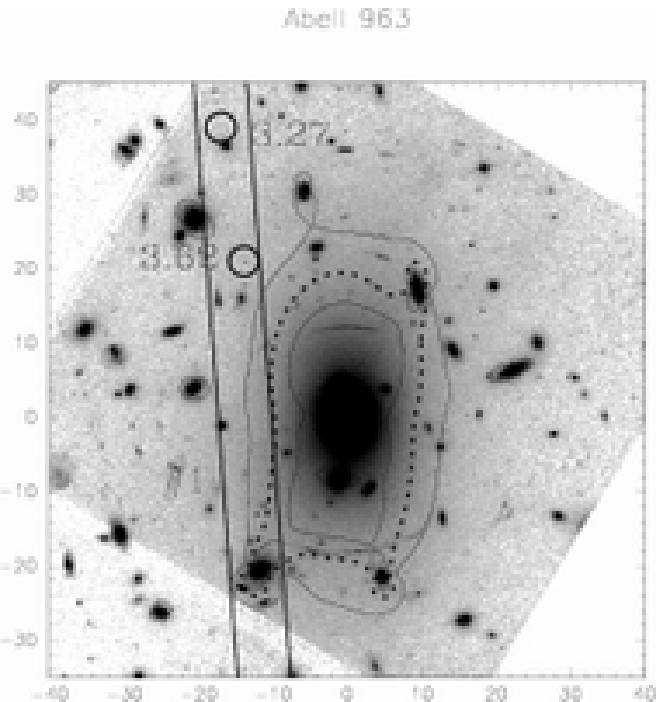}
  \includegraphics[width=7cm]{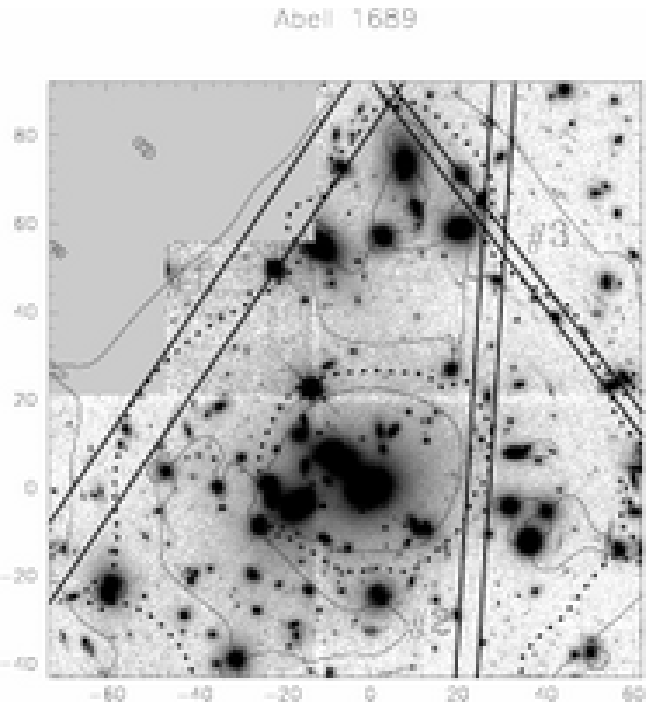}\includegraphics[width=7cm]{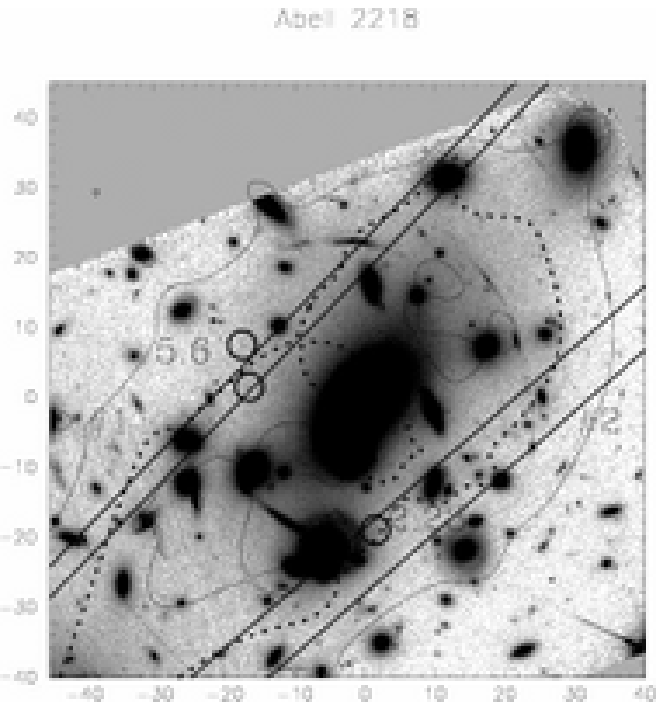}
  \caption[Survey clusters, with survey area and lensing critical
  curves.]{Survey geometry for a selected sample of our clusters.  For
  each cluster, superposed on the \textit{HST}/WFPC2 image are the
  critical lines for a source at $z=5$ (dotted lines).  The solid
  curves bound areas where the magnification for such a source exceeds
  a factor of 10.  The regions bounded by parallel straight lines are
  the long-slit survey area.  Numeric labels correspond to the key in
  Table~\ref{tab:obs}.  The axes are labeled in arcseconds. (upper
  left) Abell 68.  (upper right) Abell 370.  (middle left) Abell 773.
  (middle right) Abell 963.  (lower left) Abell 1689.  (lower right)
  Abell 2218.}
  \label{fig:johan}
\end{figure}

Given that high-quality lens models are only available for about a
dozen clusters observable from Hawaii, the total area available to us
for a survey is currently quite small.  To make the most of this
limited resource, we conducted a spectroscopic survey.  The advantage
of a spectroscopic survey is that \lya\ emission can be simultaneously
surveyed for over a redshift range 2.2$<z<$6.7.  The primary draw-back
of a spectroscopic survey was that, with the instrumentation
available, the most efficient technique was slit spectroscopy.  The
geometry of a long slit is not well matched to the lensed region of
sky (see Fig.~\ref{fig:johan}), thus some of the slit area covers area
outside the cluster that is not strongly magnified.

A long-slit survey does have other advantages. We can expect many
emission-line sources in our survey other than \lya; in particular,
optical lines associated with strong star formation, such as
[\ion{O}{2}] 3726, 3729~\AA; H$\beta$ 4861~\AA; [\ion{O}{3}] 4959,
5007~\AA; and H$\alpha$ 6563~\AA.  Low-resolution spectroscopy with
large wavelength coverage allows the rejection of many potential
low-redshift contaminants through the identification of other emission
lines.  However, the [\ion{O}{2}] doublet can be difficult to resolve
at low dispersion and, if redshifted to $z \sim 1$, there are often no
other strong emission lines present in the optical spectrum.  Thus
final identification of an emission line as \lya\ may require
follow-up spectroscopy at intermediate resolution.

Redshift identification is aided by two additional factors.  Deep
optical broadband imaging, available in at least one band for
well-studied clusters, can be used, as in narrow-band searches, as a
rejection filter: if a putative \lya\ system shows much observable
flux shortward of \lya, then it is not likely a correct line
identification, because the intrinsic UV spectrum combined with IGM
absorption create a strong decrement across the \lya\ emission line
\citep[e.g.,][]{son02}.  The second tool is available when two or more
images (due to strong lensing) of the same high-redshift source are
discovered.  In this case the lensing model itself may place a
reasonably strong constraint on the redshift of the system based on
the observed image positions and flux ratios \citep{kne96}.

The deepest survey for a given observing time would be to devote all
of the time to a single slit position. However, we expect \lya\
sources to be clustered, resulting in a non-Poisson distribution. To
estimate an accurate luminosity function we surveyed several
independent volumes (via surveying behind several lensing clusters) to
ameliorate cosmic variance and recover the Poisson noise limit (see
Section~\ref{sec:halo}).

\subsection{Survey Parameters}

Table~\ref{tab:clusters} summarizes the 9 lensing clusters of our
survey.  We have constructed a detailed mass model for each, based on
\textit{HST} imaging and lensed arc redshifts.

Details of our spectroscopic observations are listed in
Table~\ref{tab:obs}.  Clusters that were observed at multiple position
angles are designated further by an identification number.  We used
the double-beam Low Resolution Imaging Spectrograph
\citep[LRIS,][]{oke95} in long-slit mode on the Keck I 10-meter
telescope at Mauna Kea to perform our survey.  For the 2000 March
observations we used a slit 0.7\arcsec\ wide and a spectroscopic range
of $\lambda\lambda$~6800--9500~\AA\ (corresponding to \lya\ with
4.6$<z<$6.8), using a 600-line grating blazed at $\lambda$7500~\AA\
which gave a resolution of $\simeq$3.0~\AA.  In 2001 April we switched
to a 1.0\arcsec-wide slit and used a 600-line grating blazed at
1~$\mu$m over the same wavelength range as above, at a resolution of
$\simeq$4.0~\AA.  Starting in 2001 April we also began using a 300
line grism blazed at 5000~\AA\ and a dichroic at 6800~\AA\ to
simultaneously take spectra on the blue arm of the spectrograph, over
$\lambda\lambda$ 4000--6700~\AA\ (corresponding to \lya\ from
2.2$<z<$4.5) at 3.5--4~\AA\ resolution.

The length of the spectrographic slit was 175\arcsec.  We mapped an
area on the sky via offsetting the telescope perpendicular to the long
axis of the slit by a distance equal to the slit width.  At each slit
position we made two exposures of 1000~sec to facilitate cosmic ray
rejection.  Each map comprised 5--10 adjacent slit positions at the
same position angle, giving contiguous survey areas on the sky of
875--1750~arcsec$^2$.

The pointing of the slit on the sky was verified by registration of
images from the LRIS slit-viewing guide camera to the \textit{HST}
images of the cluster (because the clusters are rich in bright
galaxies, there were always many sources in the slit-viewing guide
camera images).  Our sequence of slit offsets typically agreed with a
regular spacing of the slit width to a precision of 0.1\arcsec\ (10\%
of the 1\arcsec\ slit).  More importantly, registration of the slit
position on the \textit{HST} image enabled us to look for a broadband
counterpart at the location of emission lines detected in our spectra.

The areas mapped by the procedure above were chosen to take advantage
of the strong magnification of background sources provided by the
foreground cluster.  The magnification map of the sky around the
cluster is constrained by the distribution of visible light in the
cluster and the measured velocity dispersion of some cluster members,
but is crucially verified and refined by including information from
the locations and redshifts of strongly lensed sources.  These
background galaxies, generally brighter and at lower redshift than the
$z>4.5$ sources we searched for, have been the target of previous
observations \citep[e.g.,][]{kne96}.

We used up-to-date cluster mass models to generate redshift-dependent
maps of the magnification toward $4<z<7$ sources using the
\texttt{LENSTOOL} software developed by \citet{kne93}.  The angular
diameter distance between the cluster and those redshifts depends only
weakly on redshift, so we were able to choose areas on the sky with
high magnification over our entire redshift range of interest.

The geometry of the magnification map is generally characterized by
two concentric ring-like curves of formally infinite magnification,
called the inner and outer critical lines.  These are related to the
location in the image plane of the caustic of an elliptical potential
\citep{bla92}, but modified by the deviations of the projected lensing
potential from an ellipse.  The areas of highest magnification are
found next to the critical curves, so our survey maps generally follow
the outer critical line.  The outer critical line is more amenable to
long-slit mapping because of its greater length on the sky and its
less curved form.  However, in one case (Abell 1689) we mapped sky
near the inner critical line as well.  Lensed sources close to the
critical line are often multiply imaged, forming a merging pair on
either side of the critical line.  We considered this when we chose
our map locations, but the irregular shape of the critical line,
compared to the straight shape of our slit, limited the extent to
which we could map exclusively one side of the critical line.

Figure~\ref{fig:johan} summarizes the adopted strategy for each
cluster in the context of the location of the critical line for a
lensed source at $z=5$ (dotted lines).  In the most massive clusters
with the best mass models, such as Abell~1689 and Abell~2218, we made
multiple maps (see also Table~\ref{tab:obs}).  In these cases each
survey region is labeled by a number corresponding to the observations
listed in Table~\ref{tab:obs}.

The total area on the sky covered by our survey was 4.2~arcmin$^{2}$.
The effective areal coverage of the survey is smaller due to lensing
by a spatially variable magnification factor (see
Section~\ref{sec:mag}).

\subsection{Candidate Selection and Catalog}

The 2-D spectroscopic data were reduced using standard techniques in
the NOAO/IRAF software environment\footnote{IRAF is distributed by the
National Optical Astronomy Observatories, which are operated by the
Association of Universities for Research in Astronomy, Inc., under
cooperative agreement with the National Science Foundation.}.  Cosmic
rays were rejected from each pair of images at a given location with
the \texttt{L.A.COSMIC} routine \citep{vand01} and sky emission was
removed by subtracting block-filtered data.  Sky subtraction was not
photon-limited on the strong night sky lines due to the presence of
fringing features; we account for this when determining our
sensitivity in Section~\ref{sec:suranal}.  We calibrated our absolute
efficiency with observations of spectrophotometric stars
\citep{mas90}.

The sky-subtracted 2-D spectral images were independently inspected by
two of the authors (RSE and JR) and a catalog of 46 candidate \lya\
emission lines was compiled.  Astrometric positions were determined
for each and the \textit{HST} images inspected for sources at the
relevant location. In some cases, candidates were located beyond the
boundary of the \textit{HST} images and ground-based images were used.

Candidate \lya\ emission lines were characterized on the basis of
several criteria.  First, the full spectrum (generally 4000--9500~\AA)
was closely examined for other emission lines.  On a second pass,
candidates were ranked as marginal or promising depending on their
apparent strength and spatial extent.  Out of the initial list of 46,
7 sources are confirmed \lya\ lines at $2.8<z<5.7$, and 4 sources are
promising candidates that we identify as likely to be \lya\ lines.
Those 11 sources are listed in Table~\ref{tab:cand}, with observed
line fluxes for detections at $z>4.5$ (on the red arm of the
spectrograph).

\subsection{Intermediate-resolution Spectroscopy}

The [\ion{O}{2}]~3727~\AA\ doublet has a rest-frame separation of
3~\AA.  Thus at $z\sim1$, when the doublet is redshifted into our most
important spectral range, the observed doublet separation is
$\sim6$~\AA.  This is close to our LRIS spectral resolution of 4~\AA,
so to determine whether any of our \lya\ line candidates were
unresolved [\ion{O}{2}] doublets, we followed up 15 candidates with
the Echellette Spectrograph and Imager \citep[ESI,][]{she02}.  We took
spectra using the echelle mode and a 0.75~\arcsec\ slit, which
delivered a spectral resolution of $R=6000$.  The exposure times
varied depending on candidate strength.  ESI spectroscopy confirmed
three candidates as \lya\ emission lines (see Table~\ref{tab:cand}),
at $z=3.27$, $z=3.62$, and the $z=5.57$ galaxy presented in
\citet{ell01}.  Additionally, several of the original candidates
turned out to be [\ion{O}{2}]~3727~\AA\ at $z\simeq1$.

\section{Detections}
\label{sec:detect}

\subsection{$z>4.5$}

\begin{figure}
  \includegraphics[width=20cm,angle=90]{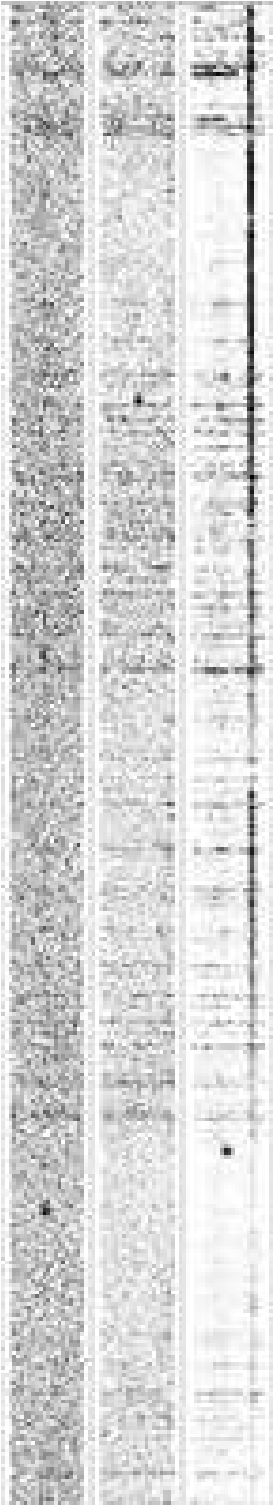}
  \caption[Two-dimensional spectra of the three confirmed $z>4.5$
  galaxies detected in the survey.]{Two-dimensional spectra of the
  three confirmed $z>4.5$ galaxies detected in the survey.  The
  spectra, from left to right, are the $z=4.89$ source in Abell 1689,
  the $z=5.57$ source in Abell 2218, and the $z=4.92$ source in
  Cl1358.  The wavelength coverage in all the spectra is 6800 to
  8430~\AA, increasing bottom to top.}
  \label{fig:three}
\end{figure}

\begin{figure}
  \includegraphics[width=3.5cm]{f4a.eps}\includegraphics[width=3.5cm]{f4b.eps}\\
  \includegraphics[width=3.5cm]{f4c.eps}\includegraphics[width=3.5cm]{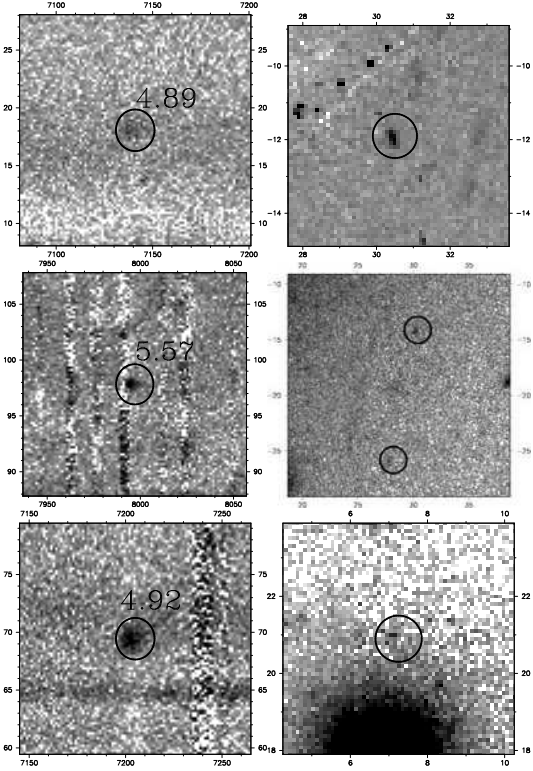}\\
  \includegraphics[width=3.5cm]{f4e.eps}\includegraphics[width=3.5cm]{f4f.eps}
  \caption[Two-dimensional spectra and \textit{HST} images for our
  three confirmed \lya\ sources.]{Two-dimensional spectra (left) and
  \textit{HST} images (right) for our three confirmed \lya\ sources.
  In the spectra the horizontal axes are labeled in \AA, and the
  vertical axes in arcsec; the image axes are labelled in arcsec.
  (upper) Abell 1689.2.f.  (middle) Abell 2218.1.a2.  (lower)
  Cl1358.1.ef.}
  \label{fig:stamp1}
\end{figure}

We detected three convincing \lya\ sources with $z>4.5$ (see
Table.~\ref{tab:cand}); each was detected in photometric conditions.
Figure~\ref{fig:three} shows their two-dimensional spectra, and
Fig.~\ref{fig:stamp1} shows a magnified view of the \lya\ emission
line as well as \textit{HST} images of the source locations.

We discovered a source toward Abell 2218 at $z=5.6$ that we discussed
in detail in \citet{ell01}.  That source is strongly magnified (a
factor of 33) and multiply imaged: we used the \textit{HST} data to
locate a second image outside of our survey region.  The redshift
identification was confirmed by an intermediate-resolution spectrum of
both images, showing them to be identical, with P-Cygni line profiles
characteristic of Lyman $\alpha$, and certainly not the [\ion{O}{2}]
3727~\AA\ doublet.  Our lensing model additionally constrained the
redshift of the source to be consistent only with the identification
of the line as Lyman $\alpha$.  The unlensed luminosity in the
\lya\ line is $(7.8\pm0.8)\times10^{41}~\mathrm{erg~s^{-1}}$.

We blindly recovered a $z=4.89$ galaxy in Abell 1689 that was
discovered originally by \citet*{fry02}.  This object is multiply
imaged, and the \citet{fry02} spectrum shows a strong break across the
line and metal absorption lines in the continuum redward of \lya,
confirming the redshift.  We estimate the magnification of this source
at a factor of 7.2, within the range of 3--14 suggested by the coarser
modeling of \citet{fry02}.  The unlensed luminosity in the \lya\ line
is $(7.4\pm0.7)\times10^{42}~\mathrm{erg~s^{-1}}$.

In the field of Cl1358, we discovered a source at $z=4.92$, the same
redshift as that of the strongly lensed arc discovered by
\citet{fra97}.  We believe our source is likely associated, as an
additional component, with that responsible for the giant arc.  The
magnification is $\times10$, giving an unlensed luminosity in the
\lya\ line of $(2.5\pm0.3)\times10^{42}~\mathrm{erg~s^{-1}}$.

\begin{figure}
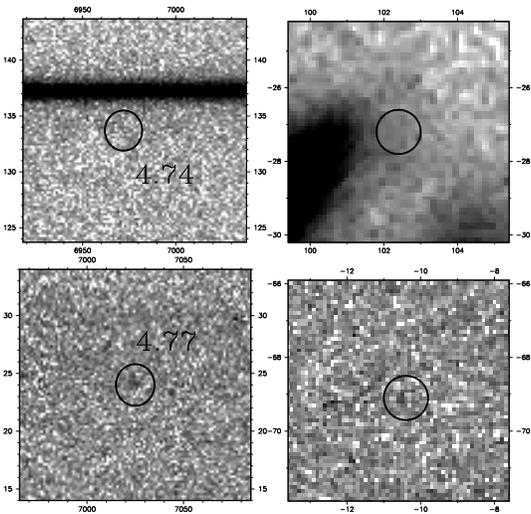

  \includegraphics[width=3.5cm]{f5a.eps}\includegraphics[width=3.5cm]{f5b.eps}\\
  \includegraphics[width=3.5cm]{f5c.eps}\includegraphics[width=3.5cm]{f5d.eps}\\
  \caption[Two-dimensional spectra and \textit{HST} images for our two
  likely \lya\ sources.]{Two-dimensional spectra (left) and
  \textit{HST} images (right) for our two likely \lya\ sources.  In
  the spectra the horizontal axes are labeled in \AA, and the vertical
  axes in arcsec; the image axes are labelled in arcsec.  (upper)
  Abell 773.1.e.  (lower) Abell 963.1.d.}
  \label{fig:stamp2}
\end{figure}

In addition to these confirmed sources, we discovered two more sources
that we consider likely to be \lya\ emission lines at $z>4.5$ (see
Table.~\ref{tab:cand}).  We discovered a source in the field of Abell
773 that we consider to be a good candidate for \lya\ emission at
$z=4.74$.  If this is the correct identification, the unlensed
luminosity in the Lyman $\alpha$ line is
$(2.8\pm0.6)\times10^{41}~\mathrm{erg~s^{-1}}$, using a magnification
factor of 9.5.  The other likely source was discovered in
non-photometric observations; though we lack absolute flux calibration
for this source, we applied a systematic photometric correction that
we consider uncertain up to a factor of approximately 2.  The likely
$z=4.77$ source is in the field of Abell 963, magnified by 2.2 times.
The source would then have an unlensed \lya\ line luminosity of
$(1.4\pm0.2)\times10^{42}~\mathrm{erg~s^{-1}}$ (statistical error
only).  Two-dimensional spectra of the emission lines and images of
the likely candidates are provided in Fig.~\ref{fig:stamp2}.

\subsection{$z<4.5$}
 
We detected with certainty four \lya\ emission-line sources at
$z<4.5$.  One was a blind recovery of a $z=2.80$ source behind Abell
370 discovered by \citet{ivi02}, and a second is another galaxy also
at $z=2.80$ in the same field.  Two are new detections, at $z=3.27$
and $z=3.62$, both behind Abell 963.  Two more good candidate \lya\
emission-line source identifications are pending.  In addition two
galaxies behind Abell 2218, both at $z\simeq2.5$, were detected in
\lya\ absorption.  We measured secure redshifts for 51 other sources
using other emission lines, primarily [\ion{O}{2}], [\ion{O}{3}], and
H$\alpha$.  These data are useful for further constraining the cluster
mass models, and will be presented in a separate paper (Richard et
al., in prep.).

\section{Survey Analysis}
\label{sec:suranal}

In this section we compute the number density of \lya\ emission-line
galaxies in our survey as a function of \lya\ line luminosity, $L$.
To accomplish this we first determine the effective volume of the
survey as a function of the luminosity and redshift of a source.

\subsection{Survey Volume}
\label{sec:survol}

A location in our survey volume is characterized both by location on
the sky, $\sky$, and a redshift, $z$.  The differential volume
element located at position ($\sky$,$z$) in our survey is
\begin{equation}
\dd V_\mathrm{c}(\sky,z) = \frac{1}{\mg}
\left[ \frac{\dd l_\mathrm{c}(z)}{\dd z} \dd z \right]
\times \left[ D_\mathrm{c}^2(z) \dd\sky \right].
\end{equation}
The first factor corrects for the lensing effect, which decreases the
area surveyed.  The second factor is the comoving length of the volume
element along the line of sight, with
\begin{equation}
\frac{\dd l_\mathrm{c}(z)}{\dd z} = 
\frac{c}{H_0 \left[\Omega_m(1+z)^3+\Omega_\Lambda\right]^{3/2}}
\end{equation}
(we have assumed a flat universe).  The third factor is the comoving
transverse area of the volume element, with
\begin{equation}
D_\mathrm{c}(z) = \int_0^z \frac{\dd l_\mathrm{c}(z')}{\dd z'} \dd z'
\end{equation}
(subscript ``c'' denotes that the quantity is measured in comoving
coordinates, which we use throughout).

Every volume element of our survey is characterized by a limiting
\lya\ line luminosity, $\llim$; a source with \lya\ line luminosity
$L$ will be detected in our survey provided it resides in a volume
element with $\llim \leq L$.  The limiting luminosity of a volume
element depends on the magnification (due to lensing by the foreground
cluster), $\mg$, the limiting observed \lya\ line flux $\flim$, and a
slit transmission function, $\slit$,
\begin{equation}
\llim = \frac{4 \pi (1+z)^2 D_\mathrm{c}^2(z)}{\slit} \frac{\flim}{\mg}.
\end{equation}

\subsubsection{Magnification, $\mg$}
\label{sec:mag}

The magnification due to lensing by a given cluster is a function of
both position and redshift.  The references for the cluster mass
models are given in Table~\ref{tab:clusters}.  These models were run
through the \texttt{LENSTOOL} software \citep{kne93} to generate the
magnification as a function of redshift at every position in the
survey.  In practice the area of the survey was divided into parcels
of sky of length 0.8\arcsec\ and width equal to the slit width, and
the magnification was calculated at the center of each parcel.  The
magnification at each position was sampled at nine redshifts, and the
magnification at a particular redshift found by interpolation.

\begin{figure}
  \includegraphics[width=8.4cm]{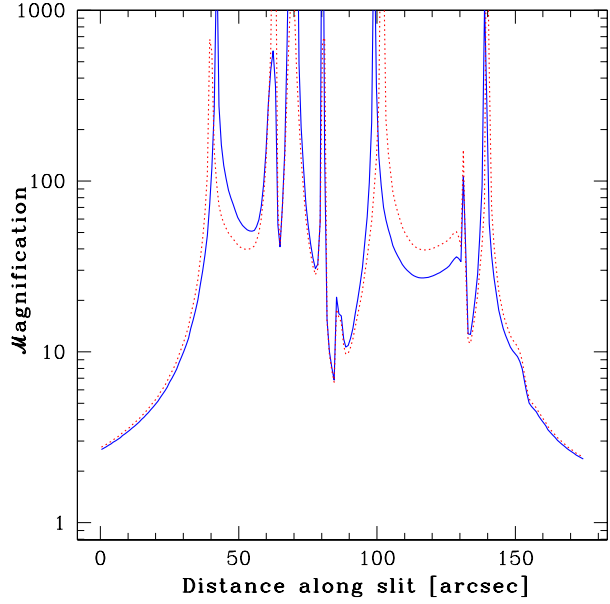}
  \caption[Magnification due to the cluster Abell 2218 of background
  sources, as a function of position and redshift.]{Magnification due to the cluster Abell 2218 of background
  sources, as a function of position and redshift.  This figure shows
  the magnification factor at positions along a 175\arcsec\ longslit
  at one of our survey positions in Abell 2218 (within pointing
  ``\#1'' in Fig. \ref{fig:johan}).  The solid curve is for sources
  at $z=4.3$, and the dotted curve is for sources at $z=6.8$.}
  \label{fig:slitmag}
\end{figure}

Figure~\ref{fig:slitmag} shows the magnification as a function of
position along a slit observed in Abell 2218, for two different
redshifts.  The magnification at a given position is a weak function
of redshift for magnification values less than $\sim100$ (94\% of the
survey area), because the angular diameter distance between the
cluster and the source changes by less than 25\% over the source
redshift range $4.5<z<6.7$.  Very near the critical lines
magnification is a stronger function of redshift.  Our survey maps sky
by observing adjacent slit positions, so errors associated with
interpolating the highest magnification $\mg>100$ points should not be
important in our estimate of the survey volume.

\begin{figure}
  \includegraphics[width=8.4cm]{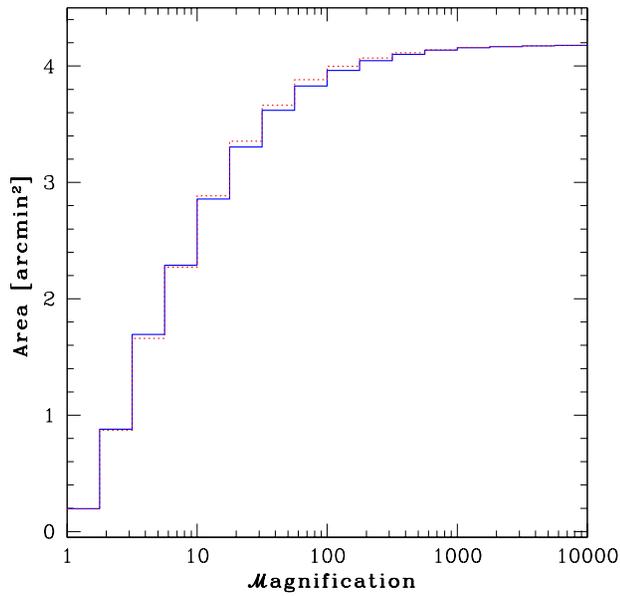}
  \caption[Cumulative histogram of the magnification factor over the
  entire survey area.]{Cumulative histogram of the magnification
  factor over the entire survey area.  The solid and dotted curves
  show the magnification histograms for sources at $z=4.3$ and
  $z=6.8$, respectively.}
  \label{fig:maghist}
\end{figure}

Figure \ref{fig:maghist} is a cumulative histogram of the
magnification factor over the survey; the area surveyed as a function
of magnification is very weak function of redshift, even at the
highest magnifications.  About half of the area we surveyed is
magnified by at least a factor of 10, with the lower magnification
values coming from area at the ends of the slits, because most
strong-lensing clusters subtend a size less than the slit length
(175\arcsec) on the sky (see Fig.~\ref{fig:johan}).

\subsubsection{Limiting \lya\ line flux, $\flim$}

We define our limiting \lya\ line flux as the signal in an aperture of
1.3\arcsec\ by 7.7~\AA\ that exceeds 5 times the root-mean-square
fluctuations (noise) in apertures of that size, i.e., a 5-$\sigma$
limit.  The spatial dimension of the aperture was chosen to be roughly
matched to the seeing, and the spectral dimension was chosen match the
expected line-width of high-redshift \lya\ emission from galaxies,
$\sim300~\kms$.  If a source is larger than our aperture, which is
especially possible along the spatial direction if the source is
strongly lensed, then we will not be as sensitive to that source as we
estimate.

We assumed that the sky noise was constant over the length of the slit
at fixed wavelength.  This allowed us to include the non-Poisson
contribution to the noise level from fringing features, which
dominated the noise on strong sky lines.  All three of our confirmed
$z>4.5$ detections were more than 5-$\sigma$ detections, but we found
that visual inspection generated candidates (some of which were
subsequently confirmed as bona fide emission lines) with fluxes below
the 5-$\sigma$ limit; in particular, one of the likely candidates at
$z>4.5$ is just at the 5-$\sigma$ limit.  Thus a 5-$\sigma$ limit
should be appropriate for the calculation of our survey volume.

The limiting line flux varies as a function of wavelength due to the
wavelength dependence of the atmospheric absorption and the
sensitivities of the telescope and instrument, but the largest
dependence is due to the strong variation in atmospheric emission from
OH airglow lines (except at $\lambda\ga9300$~\AA, where the sharp drop
in instrumental sensitivity dominates).  Figure \ref{fig:flim1} shows
$\flim$ for the slit pointing illustrated in Fig.~\ref{fig:slitmag}, a
2000~sec observation under photometric conditions.  We compute $\flim$
by simple conversion of the observed wavelength into the corresponding
redshift for \lya\ to be observed at that wavelength.

Approximately half of our survey data were taken in non-photometric
conditions.  We account for this by dividing the limiting line flux
measured from the observations by our best estimate of the sky
transparency during the exposure.  During some exposures we have
sequences of guide-camera observations that were used to measure
relative transparency between observations, and in some cases absolute
transparency when photometric guide-camera images were available.  In
other cases we rely on observation log notes based on the count rate
of the guide star as reported by the telescope operator.

\subsubsection{Slit transmission, $\slit$}

The slit widths used in our survey, originally 0.7\arcsec\ and later
1\arcsec, are comparable in size to the seeing disk.  Consequently the
transverse distance of a source from the center-line of the slit has a
small impact on the source's observability: objects in the center of
the slit are easier to detect than objects at the slit edge.  Since
the absolute calibration was performed with respect to standard stars
in the center of the slit, we compute the fraction of light
transmitted through the slit as a function of transverse position on
the slit (ignoring objects outside of the slit, as they will in
general fall on another slit), with respect to an object at the center
of the slit,
\begin{equation}
\slit = 
\frac{\mathrm{erf}\left[\frac{w+2x}{s} \beta\right] + 
\mathrm{erf}\left[\frac{w-2x}{s} \beta\right]}
{2~\mathrm{erf}\left[\frac{w}{s_0} \beta\right]}.
\end{equation}
Here $w$ is the slit width, $s$ is the seeing full-width at half
maximum (FWHM) during survey observations, $s_0$ is the seeing FWHM
during standard star observations, $x(\sky)$ is the transverse
distance of the source from the center of the slit, and
$\beta\equiv\sqrt{\ln(2)}$.  The minimum value of $\slit$ in our survey is
about 0.8, so it has a minor effect on the computation of $\llim$.

\subsection{Volume as a function of source redshift and luminosity}

The total volume of our survey sensitive to a source of \lya\ line
luminosity $L$ is the integral over all volume elements in the survey
with $\llim\leq L$,
\begin{equation}
V_\mathrm{c}(L) = 
\int_\sky \int_z \dd V_\mathrm{c}(\sky,z)~H[ L-\llim ],
\end{equation}
where $H(y)$ is the step function defined with $H(y\geq0)=1$.

\begin{figure}
  \includegraphics[width=8.4cm]{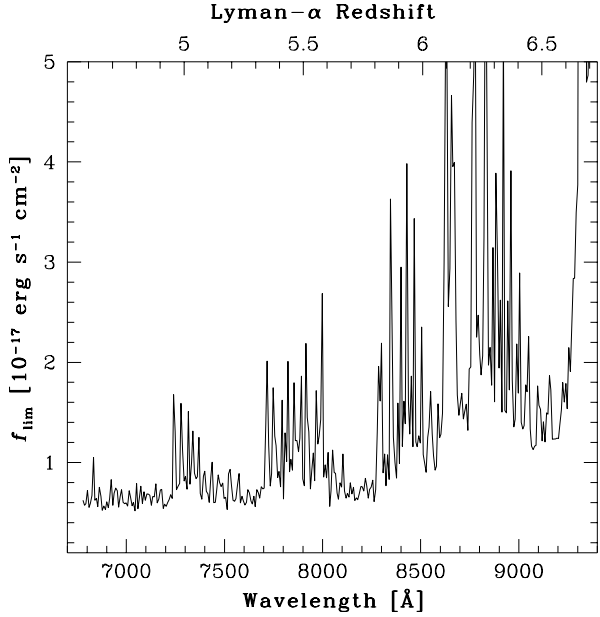}
  \caption[Limiting \lya\ line flux as a function of
  wavelength.]{Limiting \lya\ flux as a function of wavelength for the
  survey exposure time of 2000~s, assuming photometric conditions.
  The curve is the 5-$\sigma$ flux limit to detect an emission line in
  a 1.3\arcsec\ by 7.7~\AA\ aperture at that wavelength.  The top axis
  is labeled with the redshift corresponding to observed \lya\ falling
  at the wavelength on the bottom axis.  The strong fluctuations in
  $\flim$ are caused by atmospheric emission lines, and the rise at
  high redshift is due to decreased instrumental sensitivity.}
  \label{fig:flim1}
\end{figure}

\begin{figure}
  \includegraphics[width=8.4cm]{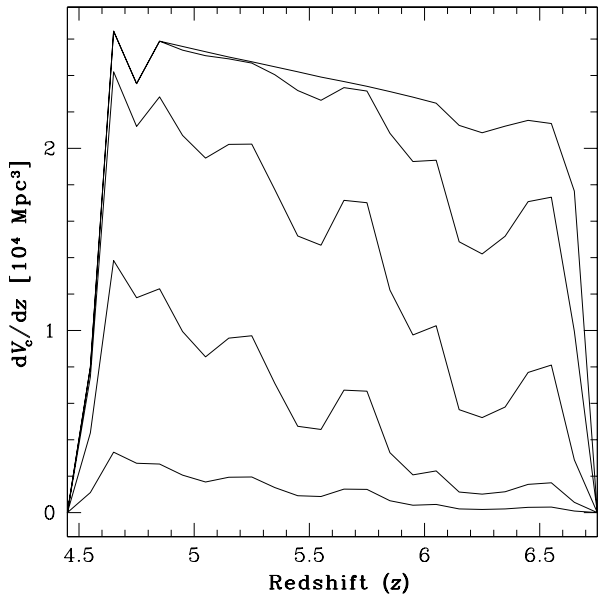}
  \caption[The redshift distribution of our survey volume.]{The
  redshift distribution of our survey volume.  The curves, from top to
  bottom, represent the redshift distribution of subsurveys down to
  limiting luminosity $\log_{10}L=(43.5,43,42.5,42,41.5)$; for yet
  lower values of the limiting luminosity, the curves have the shape
  of the bottom curve, but are scaled down (see Fig. \ref{fig:voll}).}
  \label{fig:zdist}
\end{figure}

Figure~\ref{fig:zdist} shows the redshift distribution of our survey
volume as a function of $\llim$.  The general slight decrease toward
high redshift is due to the evolution of the line element with
redshift, and the modulation is due to the wavelength-dependent
limiting line flux (see Fig. \ref{fig:flim1}).

We divide our survey arbitrarily into two redshift bins, $4.6<z<5.6$,
and $5.6<z<6.7$.  There is no natural binning choice, but by breaking
our survey at $z=5.6$ we retain almost equal survey volume (at the
brightest luminosities) in each bin.  However this places all three of
our confirmed high-redshift detections into the $4.6<z<5.6$ bin, and
none in the $5.6<z<6.7$ bin.  If we had broken the bins at $z=5.5$,
the number density in the lower redshift bin would decrease, and the
number density in the higher-redshift would increase, that is, the
removal or inclusion of a source substantially outweighs the change in
volume associated with changing the redshift binning.

\begin{figure}
  \includegraphics[width=8.4cm]{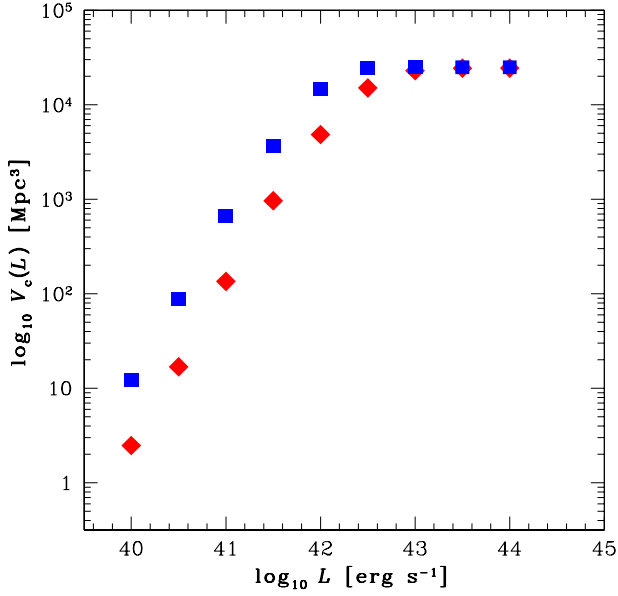}
  \caption[Survey volume sensitive to sources with \lya\ line
  luminosity $L$.]{Survey volume sensitive to sources with \lya\ line
  luminosity $L$.  The points show the volume of the survey within
  which a source of luminosity $L$ would be detected.  The survey has
  been divided into two redshift ranges: the volume with $4.6<z<5.6$
  is shown with squares, and the volume with $5.6<z<6.7$ is shown with
  diamonds.}
  \label{fig:voll}
\end{figure}

In Fig.~\ref{fig:voll} we plot the survey volume sensitive to a source
of luminosity $L$ for each of our two redshift bins (represented by
the two different symbols).  At high luminosities there is no
dependence of the survey volume on luminosity, because sources at such
high luminosities are so bright that we would detect them at any
magnification factor or redshift in our survey.  At $L=10^{42}~\ergs$
the high-redshift bin has less volume because of the stronger sky
lines at longer wavelengths (see Figs.~\ref{fig:flim1} and
\ref{fig:zdist}) and larger luminosity distance compared to the
low-redshift bin.  At lower luminosities the survey volume for both
bins falls off steadily and similarly.  This is a because only
strongly magnified volume elements contribute to the survey volume,
and the number of volume elements at a given magnification is not
sensitive to redshift (Fig.~\ref{fig:maghist}).

\subsection{Number Density}
\label{sec:numdens}

Our survey detected three confirmed and two likely $z>4.5$ sources,
so to estimate a relatively robust number density parameter, and for
comparison with other surveys, we compute a cumulative number density
of sources.  We construct the cumulative number density at each value
of the \lya\ line luminosity $L$ (in each redshift bin) by evaluating
the survey volume at that luminosity (see above), and then counting
the number of detected sources brighter than $L$ in the survey volume.

\begin{figure}
  \includegraphics[width=8.4cm]{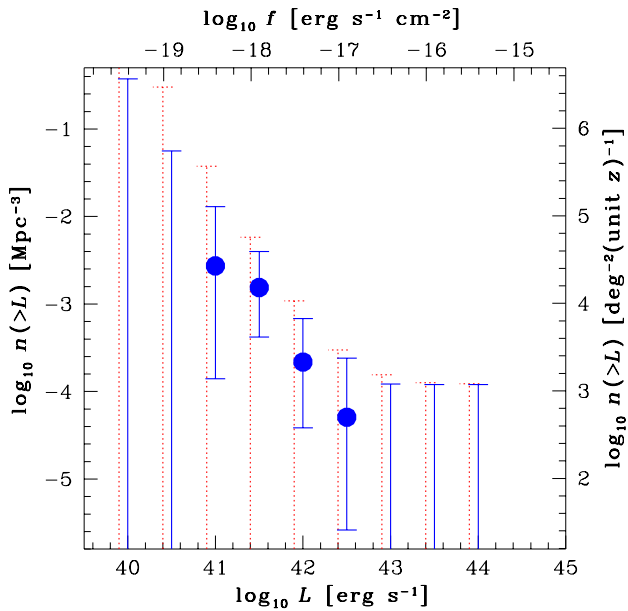}
  \caption[Number density of sources brighter than luminosity $L$, for
  the three confirmed $z>4.5$ sources.]{Number density of sources
  brighter than luminosity $L$, for the three confirmed $z>4.5$
  sources.  Each luminosity is treated as an independent sub-survey
  down to luminosity limit $L$, and the cumulative number density is
  calculated from the number of sources in the sub-survey and the
  volume of the sub-survey.  The survey was divided into two redshift
  bins: the solid lines and points are for $4.6<z<5.6$, and the dotted
  lines (off-set slightly for clarity) are for $5.6<z<6.7$.  The error
  bars are 95\% limits calculated using Poisson statistics.  The top
  and right axis are labeled in units for comparison of the
  $4.6<z<5.6$ bin results with other work: the left and bottom axes
  were transformed assuming a redshift of $z=5.0$.}
  \label{fig:numd}
\end{figure}

Figure \ref{fig:numd} shows $n(>L)$, the number density of sources
with \lya\ line luminosities greater than $L$, for our two redshift
bins, \textit{considering only the three confirmed sources}.  There
are only upper limits at $L\geq10^{43}~\ergs$ because although all
three detected sources are in the survey volume, none was that
luminous.  Our most luminous source is $7.4\times10^{42}~\ergs$, so
the first data point appears at $L=10^{42.5}~\ergs$ (in the
low-redshift bin).  All three of our confirmed detections contribute
to the $L=10^{41.5}~\ergs$ point because all are brighter than that
limit, and each would have been detected even if its luminosity were
only $10^{41.5}~\ergs$.  In contrast at $L=10^{41}~\ergs$, all three
confirmed detections are still brighter than this luminosity, but only
one is located inside the $L=10^{41}~\ergs$ survey volume.  At yet
fainter luminosities we are back to upper limits because none of the
three confirmed detections would have been discovered if it were
fainter than $10^{41}~\ergs$.

In our high-redshift bin we have no detections, and thus can provide
only upper limits at all luminosities.  It is clear that though we can
rule out a strong increase in the number counts of \lya\ emitters at
$5.6<z<6.7$ compared to $4.6<z<5.6$, we cannot further constrain the
number-count evolution.  In particular, our results are consistent
with no evolution or a decrease with increasing redshift in the \lya\
source counts as a function of redshift over $4.6<z<6.7$.

All upper limits and error bars in Fig.~\ref{fig:numd} are 95\%
confidence limits calculated using Poisson statistics.  We have
conceptually divided our survey into sub-surveys sensitive down to
different \lya\ line luminosities, but these sub-surveys are not
independent (and in fact highly correlated).  If, for example, a
theoretical curve passed just through the upper error bars of two
points, our data would indicate roughly a 95\% inconsistency, not a
99.8\% inconsistency.

The right and top axes of Fig.~\ref{fig:numd} are labeled with unit
conversions of the left and bottom axes, assuming the data fall at
$z=5$.  These serve to allow a rough reference of our results to be
easily read off in the other units commonly used to describe the
abundance of \lya\ emission-line galaxies.  The right and top axes are
inapplicable to our high-redshift bin upper limits.

Figure \ref{fig:numd} shows $n(>L)$, the number density of sources
with \lya\ line luminosities greater than $L$, for our two redshift
bins, \textit{considering all five confirmed and likely sources}.
Since we added detections while keeping the survey volume fixed, the
number densities increased.  The number densities of this sample are
still marginally consistent with the 95\% confidence limits from
Fig.~\ref{fig:numd}.

\section{Comparison with Other Observations and with Theory}
\label{sec:comp}

\subsection{Comparison with Other Observations}
\label{sec:compobs}

In Table~\ref{tab:surveys} we present parameters inferred from our
survey and existing $z\sim5$ galaxy surveys.  The first five entries
in the table describe our survey, divided by redshift bin and
sub-survey limiting \lya\ line luminosity.  The remaining surveys
above the horizontal rule are other \lya\ emission-line surveys.
Surveys below the horizontal rule are Lyman-break galaxy surveys,
described later in this section.

There are two entries in the number of sources column for each row
corresponding to our data.  The first number is the total number of
confirmed and likely \lya\ sources in that sub-survey, and the second
number, in parentheses, in the number of those that are confirmed.  We
report the corresponding number densities analogously in the density
column.

The limiting luminosity, volume, and number density of each of the
previously published \lya\ emission-line surveys appearing in
Table~\ref{tab:surveys} do not always appear in the corresponding
reference.  As necessary we have used the published information to
calculate those values ourselves (for example, converting a limiting
line flux and redshift into a limiting line luminosity).  We expect
that the final results published by the groups may differ somewhat.
In particular, there may be a publication bias toward surveys with
discoveries, so it is possible there is some bias in the data
presented toward higher number density.  We have included only
systematic \lya\ emission surveys, because reconstructing the survey
volumes of published serendipitous discoveries was not possible.

Fig.\ref{fig:tmplfmf} plots the data from the $4.6<z<5.6$ bin of our
survey (as solid circles) with the data from the other \lya\ surveys
listed in Table~\ref{tab:surveys} (open squares).  Note that these
points are in general from different redshifts, and no redshift
correction has been applied.  The error bars shown are 95\% confidence
limits assuming Poisson errors.

A comparison of our data with published results shows that, by
utilizing strong lensing, we have provided meaningful upper limits on
the population of \lya\ emission-line galaxies two orders of magnitude
fainter than previous surveys, in addition to providing confirmed data
one order of magnitude fainter.  With existing observational
facilities, lensed surveys are the only way to probe to such depth.

At $L=10^{42.5}~\ergs$, where our survey overlaps other \lya\ surveys,
there is marginal consistency between our data and published results.
Most of the other \lya\ surveys are narrow-band photometric surveys
(in particular the three surveys with $>10$ sources in
Table~\ref{tab:surveys}), where the points plotted do not represent
confirmed sources, but rather photometric candidates corrected for the
spectroscopic success rate of a small sample.  As noted previously
this figure represents data reported inhomogeneously, so some of the
scatter may be related to the different redshift ranges and \lya\
equivalent-width criteria of the surveys, as well as errors in
contamination estimation and possible errors in our interpretation of
published information.

We detected no sources at $z>5.6$.  This is marginally inconsistent
with existing data of source densities at $z=5.7$ and $z=6.5$.
However, our lack of sources at $z>5.6$ compared to other surveys is
qualitatively consistent with our smaller number density of
$4.6<z<5.6$ sources at $L=10^{42.5}~\ergs$.

For comparison, we have plotted results from four $z\sim5$ Lyman-break
galaxy (LBG) surveys \citep{sta03,yan03,iwa03,fon03,leh03}.  The
parameters for these surveys are listed in Table~\ref{tab:surveys},
below the horizontal rule.  Again we have converted published data
into number density as necessary, and taken a further step to plot
those points on a \lya\ line luminosity scale: the LBG survey limit
was converted into a rest-frame UV continuum limit, then into a
star-formation rate using the relation of \citet{ken98}, then into a
\lya\ line luminosity assuming $1~\sfr$ of star formation produces
$10^{42}~\ergs$ in the \lya\ line \citep[after converting H$\alpha$
luminosity into \lya\ luminosity]{ken98}.  No unmitigated conclusions
can be drawn from this comparison, though it is intriguing that the
$z\sim5$ LBG surveys may be discovering the same population as the
\lya\ emission-line galaxies, if the \lya\ line is typically 1/3 the
value expected based on the UV continuum SFR.  This is similar to the
ratio observed in the $z=5.7$ sample of \lya\ emitters of
\citet{aji03} and in two galaxies at $z=6.5$ by \citet{hu02b} and
\citet{kod03}.  However, four of six confirmed $4.8<z<5.8$ galaxies
selected by the Lyman-break technique by \citet{leh03} have \lya\ line
fluxes less than 10\% of the values naively predicted from their UV
continuum SFRs.

\subsection{Comparison with Theoretical Models}
\label{sec:halo}

\begin{figure}
  \includegraphics[width=8.4cm]{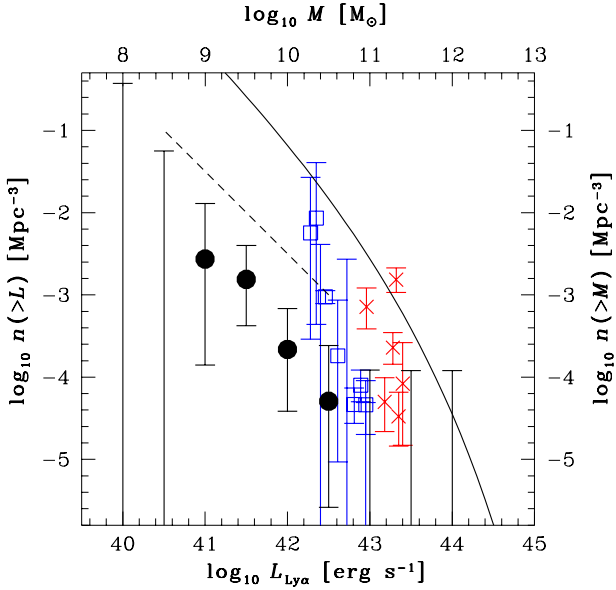}
  \caption[Number density of $z\sim5$ sources brighter than $L$, for
  \lya\ surveys and Lyman-break galaxy surveys.]{Number density of
  $z\sim5$ sources brighter than $L$, for \lya\ surveys and $z\sim5$
  Lyman-break galaxy (LBG) surveys.  The solid circles are our
  cumulative number densities of \textit{confirmed} sources brighter
  than $L$ for sub-surveys within the $4.6<z<5.6$ bin (described in
  Fig. \ref{fig:numd}).  The open squares are the cumulative number
  densities of sources brighter than $L$ inferred from other $z\sim5$
  \lya\ line surveys, and the crosses are data from $z\sim5$ LBG
  surveys.  The LBG surveys were converted to equivalent \lya\ line
  luminosities (see Section~\ref{sec:compobs}).  The long-dashed curve
  is a prediction from \citet{hai99}.  The solid curve is the
  cumulative number density of halos above the mass given on the top
  axis; the vertical scale is the same.  The data are described in
  Table~\ref{tab:cand}.}
  \label{fig:tmplfmf}
\end{figure}

First we compare our results with the \lya\ emitter model of
\citet{hai99}, who predicted the abundance of \lya\ emitters over a
range of redshifts and luminosities.  In Figs.~\ref{fig:tmplfmf} and
\ref{fig:newtmplfmf} we plot the predictions of their fiducial model
at $z=5$ as a long-dashed curve from $\log_{10}L=40.5$ to 42.5.  The
shape of the luminosity function predicted by \citet{hai99} is similar
to our observed points, but their fiducial model predicts
approximately an order of magnitude more sources than we find.  Their
models could be reconciled with our data by adopting mass-dependent
values of the star-formation efficiency or covering fraction of dusty
clouds inside the galaxies.

\begin{figure}
  \includegraphics[width=8.4cm]{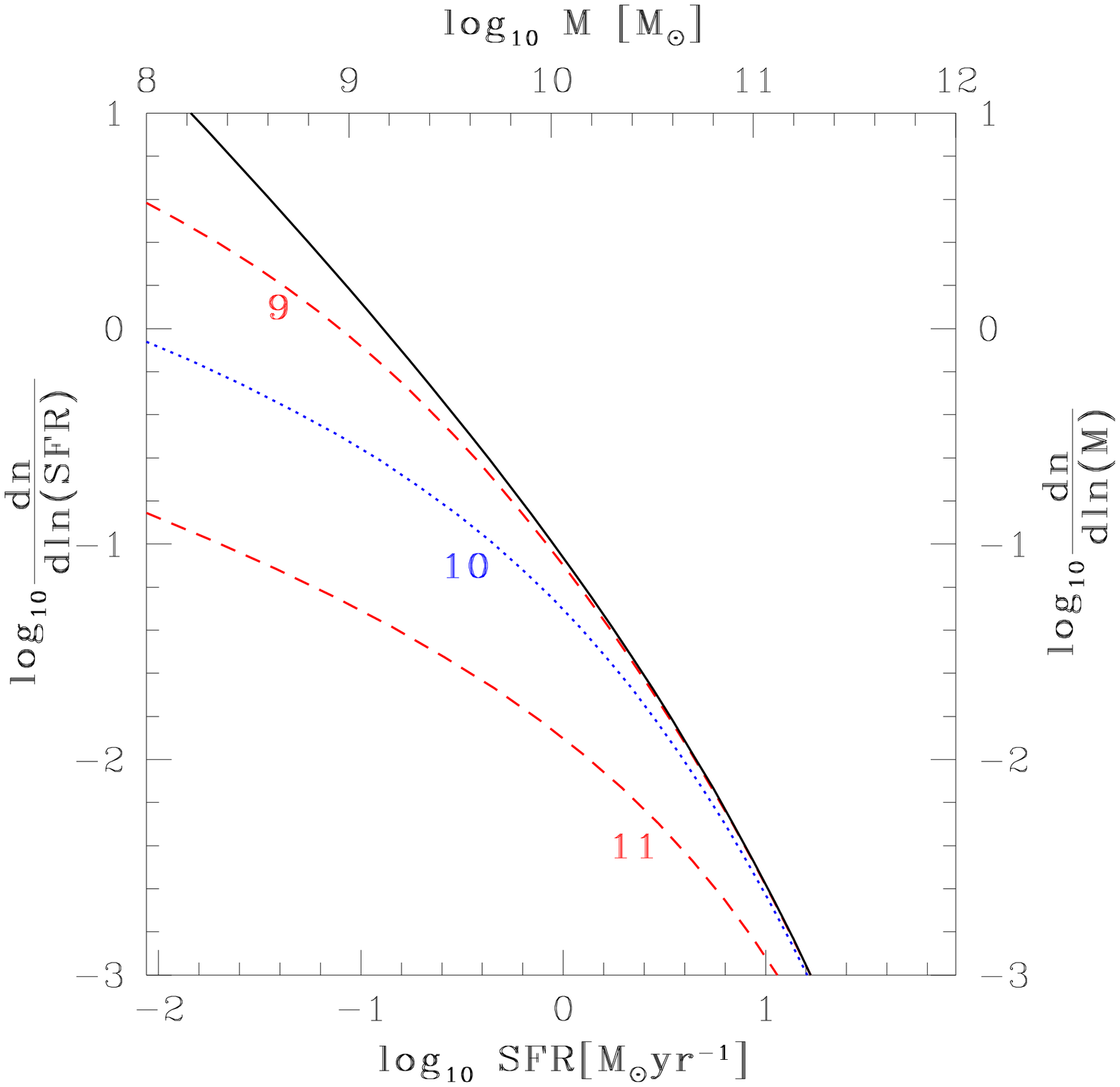}
  \caption[The star-formation rate function based on a dark matter
  halo model.]{The star-formation rate function based on a dark matter
  halo model.  The solid curve is the abundance (on the right axis) of
  halos at $z=5$ as a function of the mass on the top axis.  The
  bottom axis is a simple conversion of the halo mass into the
  expected star formation in that halo, so that the solid curve may
  also be read as a theoretical star-formation rate function using the
  bottom and left axes.  The broken curves represent cases where star
  formation in low-mass halos has been suppressed; each is labeled
  with the logarithm of the mass scale below which suppression occurs.
  See Sections~\ref{sec:lowlum} and \ref{sec:halo} for details.}
  \label{fig:schemec}
\end{figure}

As a basis for comparing our results with a simple theoretical model,
in Figs.~\ref{fig:tmplfmf} and \ref{fig:newtmplfmf} we re-plot the
luminosity function from Fig.~\ref{fig:schemec} (converted into
cumulative form), assuming no suppression of the SFR in low-mass
halos.  This simple interpretation of the \lya\ luminosity function
relates the number density of galaxies to dark matter halos.  We then
converted baryons within those dark matter halos into stars, and
stellar ionizing light into \lya\ photons.  Unlike \citet{hai99} we
made no attempt to model the radiative transfer of the \lya\ photons.

Specifically, we assumed that 10\% of the baryons in each halo were
converted into stars every halo dynamical time (defined as the ratio
of the halo virial radius to the halo circular velocity at the virial
radius).  The Hubble time at $z=5$ is roughly 10 times longer than the
halo dynamical time, thus it is possible for such halos to maintain
steady star formation at this rate.  Star-formation rate was converted
into an ionizing-photon rate using a Salpeter IMF with 1/20 solar
metallicity \citep{lei99}.  We assumed that 10\% of the ionizing
photons escape the emitting galaxy, and that 2/3 of the remaining
photons are converted into \lya\ emission.

The luminosity function predicted by this simple model provides a poor
fit to our data.  In the context of the model, it is instructive to
think of two modifications that would make the predicted luminosity
function more closely match our data.  The first is to decrease the
efficiency factors used to convert halo mass to \lya\ luminosity in a
given halo.  Alternately, the model curve could be brought into
agreement with our data if the efficiency factor for the production of
\lya\ was correct for a fraction of halos, but the rest had no
observable \lya\ emission at the time of observation.

There are three efficiencies that contribute to the overall conversion
of halo mass into \lya\ luminosity, namely the fraction of baryons
converted into stars per halo dynamical time, the stellar emissivity
of ionizing photons, and the fraction of ionizing photons observed as
\lya\ emission.  Lowering the combined efficiency by 1.5--2 orders of
magnitude would bring the model luminosity function into close
agreement with our data.  There is some difference in the shape of the
curve compared to our data, but this difference is not significant.

In the case where only a fraction of halos contain \lya\ emitters, we
would require about 1\% of halos to contain emitters at any given
time.  This could be because, in contrast to the simple model we
described, star formation is episodic in nature.  In addition, there
could be a timescale associated with the escape of \lya\ photons, such
that, for example, dust extinguishes \lya\ emission at the beginning
of a starburst, but eventually the dust is expelled and the \lya\
emission line becomes visible \citep[e.g.,][]{sha03}.  If only some
halos contain galaxies, for whatever reason, then this formalism of
assuming only a fraction of halos contain \lya\ emitters can also be
used, where the fraction now represents a filling factor, rather than
a duty cycle.

The resolution of the discrepancy between our data and the model curve
has important implications for the mass of the halos that contain the
\lya\ emitters.  If we over-estimated the \lya\ photon production
efficiency in our model, then the association between halo mass and
\lya\ emitter expressed in Figs.~\ref{fig:tmplfmf} and
\ref{fig:newtmplfmf} is not correct: the halo mass of our population
of emitters at $L\simeq10^{41.5}~\ergs$ should be $\sim10^{11}~\msun$.
This is the largest mass that could be inferred for this population,
assuming a maximum of one \lya\ emitter per halo.  From the arguments
of Section~\ref{sec:lowlum}, we may expect that halos with masses
$\ga10^{10}~\msun$ to form stars roughly similarly to one another,
i.e., though negative feedback may be important in regulating star
formation, it is ineffective in halos this massive.  Consequently, for
this low-efficiency, high halo-mass solution to the discrepancy, we
expect that our data should follow the shape of the dark-matter halo
mass function, which they do.

In contrast, if we resolve the discrepancy between the model and our
data by assuming that the efficiency we calculated is correct for a
fraction of halos, and the rest are empty of \lya\ emission, then the
mass association in Figs.~\ref{fig:tmplfmf} and \ref{fig:newtmplfmf}
is correct.  This implies a halo mass of only $\sim10^{9.5}~\msun$ for
our \lya\ emitters at $L\simeq10^{41.5}~\ergs$.  Depending on the
characteristic mass scale where negative feedback becomes a dominant
process, the \lya\ luminosity function may already deviate from the
shape of the mass function at $\sim10^{9.5}~\msun$ (see
Fig.~\ref{fig:schemec}).  Our data are slightly flatter than the mass
function, and consistent with any of the luminosity functions plotted
in Fig.~\ref{fig:schemec}.

Our theoretical interpretation so far has relied exclusively on our
data, which is consistent with the shape of the relevant halo mass
function.  However, if we consider all of the available \lya\ data,
there is some evidence for a flatter \lya\ luminosity function.  The
heterogeneous nature of the \lya\ survey data plotted in
Figs.~\ref{fig:tmplfmf} and \ref{fig:newtmplfmf} makes it difficult to
draw firm conclusions, but a combination of data at
$L\ga10^{42.5}~\ergs$ suggests that our data point at
$L=10^{41.5}~\ergs$ may be 0.5--1 dex lower than an extrapolation of
the \lya\ luminosity function from higher luminosity, assuming the
luminosity function shape matches the mass function shape.  Thus we
conclude that our data, in combination with other \lya\ surveys,
suggest that strong negative feedback is suppressing the
star-formation rate, and thus \lya\ luminosity, in our sources.

\citet{ham03} used clustering data to estimate the mass of the halos
containing \lya\ emitters at $z=4.9$.  They concluded that the
characteristic halo mass of those sources is $5\times10^{12}~\msun$
(\citealt{shi03} find a halo mass of $\sim10^{12}~\msun$ for similar
$z=4.9$ emitters on the basis of a large-scale structure feature in
their survey).  This conclusion would support the low-efficiency, high
halo-mass solution to the difference between our model luminosity
function and our data.  However, the number density of $z=4.9$
emitters is larger, by about a factor of five to ten, than the number
density of $10^{12}~\msun$ halos \citep{ham03}.  This implies,
contrary to our assumption above, that there is more than one \lya\
source per halo.  The virial radius of a $z=5$, $10^{12}~\msun$ halo
is 8.5\arcsec, so multiple sources inside a single halo may be
observed as separate sources, though this should create a very
distinct signature in the spatial distribution of sources (or extended
nature, if the sources are unresolved) that has not been reported by
other \lya\ emitter surveys.

\begin{figure}
  \includegraphics[width=8.4cm]{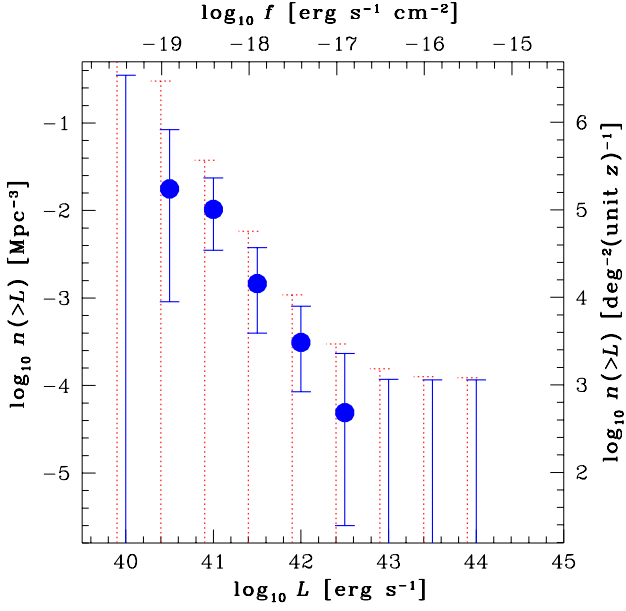}
  \caption[Number density of sources brighter than luminosity $L$, for
  the three confirmed plus two likely $z>4.5$ sources.]{Number
  density of sources brighter than luminosity $L$, for the three
  confirmed plus two likely $z>4.5$ sources.  As in
  Fig.~\ref{fig:numd}.}
  \label{fig:newnumd}
\end{figure}

\begin{figure}
  \includegraphics[width=8.4cm]{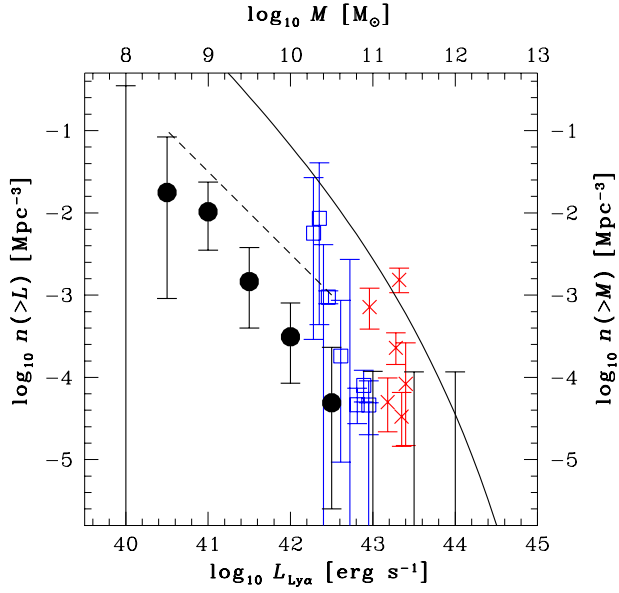}
  \caption[Number density of $z\sim5$ sources brighter than $L$, for
  \lya\ surveys and $z\sim5$ Lyman-break galaxy surveys.]{Number
  density of $z\sim5$ sources brighter than $L$, for \lya\ surveys and
  $z\sim5$ Lyman-break galaxy (LBG) surveys.  The solid circles are
  our cumulative number densities of \textit{confirmed and likely}
  sources brighter than $L$ for sub-surveys within the $4.6<z<5.6$ bin
  (described in Fig. \ref{fig:newnumd}).  The other symbols are the
  same as in Fig.~\ref{fig:tmplfmf}.}
  \label{fig:newtmplfmf}
\end{figure}

While current information on the masses of \lya\ emitter halos is
still limited, progress will continue to be made at
$L\ga10^{42.5}~\ergs$ by large \lya\ surveys.  Unfortunately, surveys
for low-luminosity \lya\ emitters will not provide sufficient survey
area for clustering studies in the near future.  Lensed surveys such
as ours, in particular, do not lend themselves to easy clustering
analysis, because the contiguous survey volume is very complex and
limited in size by the mass of the lensing foreground cluster.  As an
aside we comment that Poisson errors dominate the uncertainty in the
number densities plotted in Figs.~\ref{fig:numd} and \ref{fig:newnumd}
(and Figs.~\ref{fig:tmplfmf} and \ref{fig:newtmplfmf} for our survey),
assuming the maximum mass for the halos containing our emitters (i.e.,
every halo contains a source; see above for a caveat), and using the
clustering formalism of \citet{mo02}.

Until the advent of large-area, low-luminosity \lya\ surveys, the only
constraint on the mass of the halos containing the emitters, and thus
the only path toward understanding the suppression of star-formation
in low-mass halos, lies in detecting source populations with high
number densities, such that the halo mass function, and the assumption
that there is at most one source per halo, can be used to infer a
maximum halo mass for the population of \lya\ emitters.  This is
strong motivation for future surveys to continue to use strong lensing
to survey small volumes to considerable depths for faint, \lya\
emitting sources.

\section{Summary}
\label{sec:sum}

We performed a systematic survey for \lya\ emission at $2.2<z<6.7$
using strong lensing from intermediate-redshift clusters of galaxies
to boost our survey sensitivity to unprecedented depths.  We detected
three confirmed and two likely \lya\ emitting galaxies at
$4.7<z<5.6$, with \lya\ line luminosities of
$2.8\times10^{41}~\ergs<L<7.4\times10^{42}~\ergs$.  Our survey covered
4.2~arcmin$^2$ on the sky, with a maximum volume of
$4\times10^4~\mpc^3$ over $4.6<z<6.7$.  We find no evidence for
redshift evolution of the number density of \lya\ emitting galaxies
between $z\sim5$ and $z\sim6$, though our data are also consistent
with a decrease in number density with increasing redshift.

We present the first meaningful constraints on the the luminosity
function of \lya\ emitters at $4.6<z<5.6$ over the \lya\ luminosity
range $10^{40}~\ergs<L<10^{42}~\ergs$, corresponding to inferred
star-formation rates of 0.01--1$~\sfr$.  From a consideration of the
number density of dark-matter halos, we conclude that our population
of sources at $L\sim10^{41.5}~\ergs$ resides in halos of mass
$\la10^{11}~\msun$.

Our number density data are consistent with a \lya\ luminosity
function with the same shape as the halo mass function, but a
consideration of all available \lya\ survey data implies that we have
observed a flattening of the \lya\ luminosity function with respect to
the halo mass function.  We may have detected evidence of the
suppression of star-formation in low-mass halos at high redshift, as
predicted by theoretical models of galaxy formation.

\acknowledgements

We thank Alice Shapley for many enlightening conversations.  We thank
Graham Smith for help with mass modeling of some clusters.  We also
thank Pieter van Dokkum, Andrew Firth, and Tommaso Treu for help
obtaining and reducing the observations.

We gratefully acknowledge the helpful staff at Keck Observatory, and
the teams responsible for the creation and maintenance of the
telescopes and instruments there.

The authors recognize and acknowledge the very significant cultural
role and reverence that the summit of Mauna Kea has always had within
the indigenous Hawaiian community.  We are most fortunate to have the
opportunity to conduct observations from this mountain.

MRS acknowledges the support of NASA GSRP grant NGT5-50339.  JPK
acknowledges support from CNRS and Caltech.

\newpage

\begin{table}
\caption{Clusters surveyed}
\begin{tabular}{ccccc}
\tableline\tableline
Cluster & Redshift & RA\tablenotemark{a} & Dec\tablenotemark{b} & Lens Model Reference \\
\tableline
Abell 68 & $0.255$ & 00 36 59  & +09 09 & (1)\\

Abell 370 & $0.375$ & 02 37 18  & -01 48 & (2)\\

Abell 773 & $0.217$ & 09 14 30  & +51 55 & (1) \\

Abell 963 & $0.206$ & 10 17 09  & +39 01 & (1) \\

Abell 1689 & $0.183$ & 13 09 00  & -01 06 & (3) \\

Cl1358.1+62.45 & $0.328$ & 13 59 54  & +62 31 & (4) \\

Abell 2218 & $0.176$ & 16 35 42  & +66 19 & (5,6) \\

Abell 2219 & $0.226$ & 16 38 54  & +46 47 & (1) \\

Abell 2390 & $0.228$ & 21 53 35  & +17 40 & (7) \\

\tableline
\end{tabular}
\tablenotetext{a}{units of HH MM SS}
\tablenotetext{b}{units of +DD MM}
\tablerefs{(1) \citealt{smi03}; (2) \citealt{bez99}; (3) J. P. Kneib,
unpublished; (4) \citealt{fra97}; (5) \citealt{kne96}; (6)
\citealt{ell01}; (7) \citealt{pel99}}
\label{tab:clusters}
\end{table}

\begin{table}
\caption{LRIS Survey Observations}
\begin{tabular}{ccccc}
\tableline\tableline
Date & Cluster & Position Angle\tablenotemark{a} & 
Integration time\tablenotemark{b} &
Photometric? \\
\tableline
Mar 2000 & Abell 773     & -46.8 & 20 & Yes \\
         & Abell 1689 \#1& 84.1  & 23 & Yes \\
Apr 2001 & Abell 1689 \#2& 43    & 10 & Yes \\
         & Abell 2218 \#1& -44   & 10 & Yes \\
         & Cl1358     & -15   & 12 & Yes \\
Oct 2001 & Abell 370  \#1& -8    & 14 & Yes \\
Apr 2002 & Abell 963     & 3.6   & 14 & No \\
         & Abell 2218 \#2& -49.2 & 14 & No \\
May 2002 & Abell 1689 \#3& 12.3  & 20 & No \\
         & Abell 2219    & -69   & 14 & No \\
Sep 2002 & Abell 370  \#2& 1.7   & 14 & No \\
         & Abell 2390    & -63   & 12 & No \\
         & Abell 68      & -40   & 12 & No \\
\tableline
\end{tabular}
\tablenotetext{a}{in degrees North through East}
\tablenotetext{b}{in ksec}
\label{tab:obs}
\end{table}

\begin{figure*}
  \includegraphics[width=7cm]{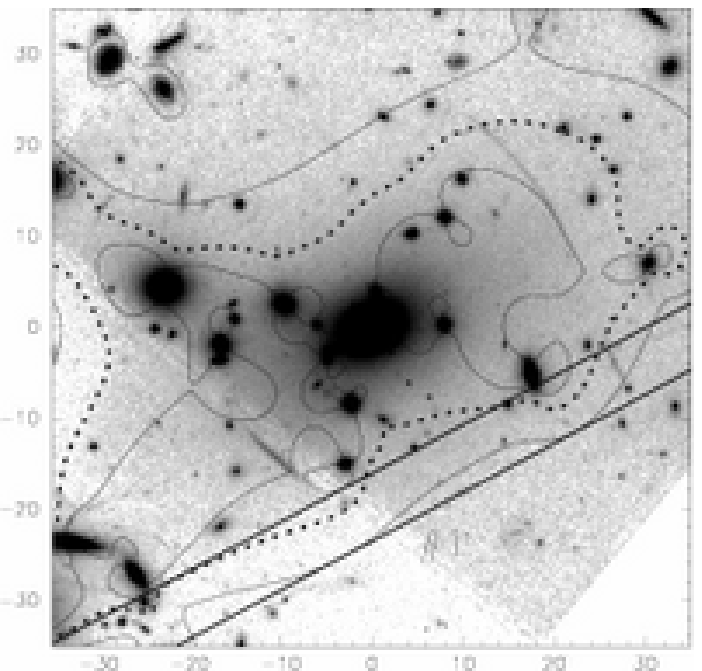}\includegraphics[width=7cm]{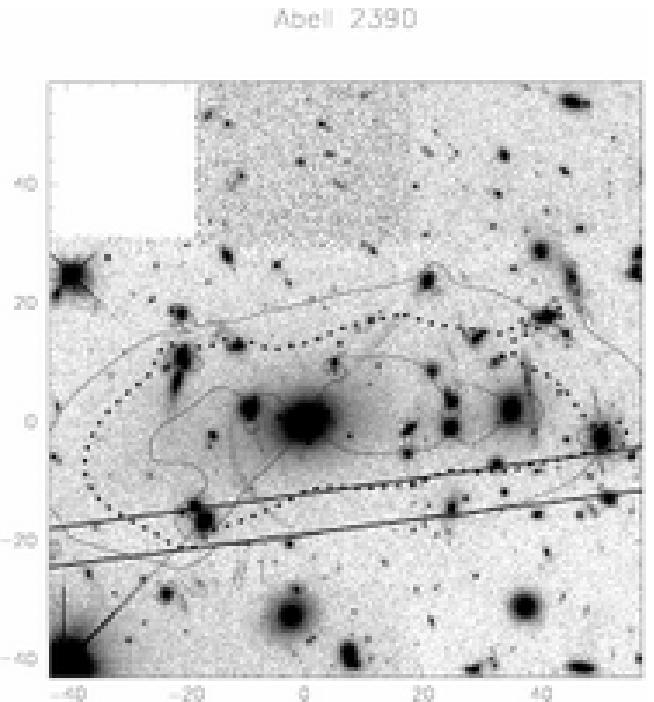}
  \includegraphics[width=7cm]{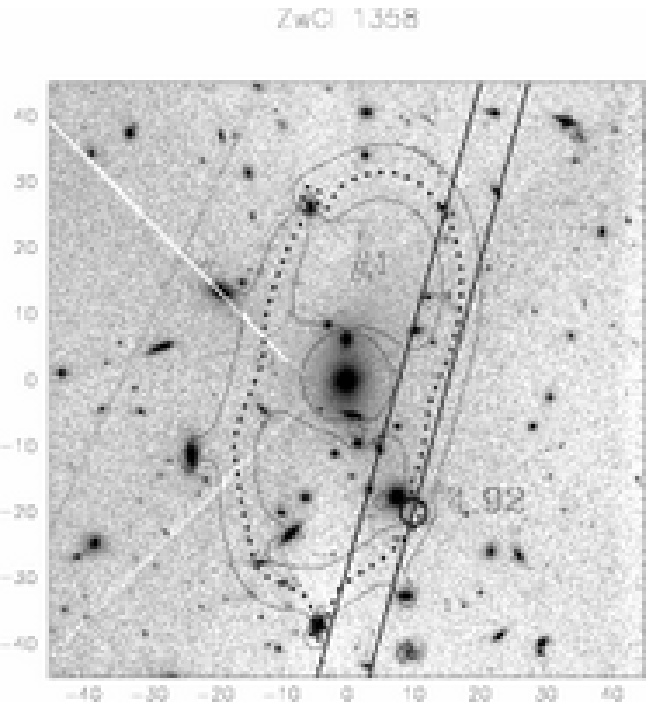}\includegraphics[width=7cm]{f2j.eps}
  \\
  \noindent{Fig. \ref{fig:johan} cont.--- (upper left) Abell 2219.
  (upper right) Abell 2390.  (lower left) Cl1358.}
\end{figure*}

\begin{table}
\caption{\lya\ Emission-Line Candidates}
\begin{tabular}{ccccccc}
\tableline\tableline
Cluster ID & RA & Dec & $\lambda$\tablenotemark{a} & $z\tablenotemark{b}$ & 
Flux\tablenotemark{c} & Comments \\
\tableline
Abell 370.1.f  &  2:39:50.60 & -1:33:45.0 & 4628 & 2.80 & & confirmed (LRIS) \\
Abell 370.1.g  &  2:39:51.80 & -1:35:57.6 & 4630 & 2.80 & & confirmed \citep{ivi02} \\
Abell 963.1.cd & 10:17:05.10 & 39:03:30.5 & 5191 & 3.27 & & confirmed (ESI) \\
Abell 963.1.efg& 10:17:04.77 & 39:03:11.0 & 5617 & 3.62 & & confirmed (ESI) \\
Abell 1689.2.f & 13:11:25.38 & -1:20:52.4 & 7141 & 4.82 & 3.0 & confirmed \citep{fry02} \\
Abell 2218.1.a2& 16:35:51.75 & 66:12:45.6 & 8001 & 5.58 & 4.4 & confirmed \citep[ESI,][]{ell01}\\
---            & 16:35:51.89 & 66:12:51.5 & & & & 2nd image \\
Cl1358.1.ef &  3:59:49.19 & 62:30:44.8 & 7205 & 4.92 & 10 & confirmed \citep{fra97} \\
Abell 773.1.e  &  9:17:55.31 & 51:44:26.6 & 6978 & 4.74 & 1.1 & likely \\
Abell 963.1.d  & 10:17:04.45 & 39:01:47.1 & 7025 & 4.77 & 0.69 & likely \\
Abell 2218.1.a1& 16:35:45.25 & 66:13:26.4 & 4216 & 2.47 & & likely \\
Abell 2218.2.b & 16:35:48.78 & 66:12:24.9 & 3928 & 2.23 & & likely \\
\tableline
\end{tabular}
\tablenotetext{a}{Wavelength of emission line in units of \AA}
\tablenotetext{b}{Source redshift assuming emission line is \lya}
\tablenotetext{c}{Observed line flux (uncorrected for lensing,
corrected for transparency) in units of $10^{-17}~\ergscm$}
\label{tab:cand}
\end{table}

\clearpage

\begin{table}
\caption{Galaxy surveys at $z\sim5$}
\begin{tabular}{cccccc}
\tableline\tableline
Redshift & log$_{10}L$\tablenotemark{a} & Number\tablenotemark{b} & Volume\tablenotemark{c} & Density\tablenotemark{d} & Reference\\
\tableline
4.6-5.6 & 40.5  & 1(0)  & 0.0053 & $1.8\times10^{-2}(0)$ & (1)\\
4.6-5.6 & 41    & 4(1)  & 0.037 & $1.1\times10^{-2}(2.7\times10^{-3})$ & (1)\\
4.6-5.6 & 41.5  & 3(3)  & 0.20  & $1.5\times10^{-3}(1.5\times10^{-3})$ & (1)\\
4.6-5.6 & 42    & 3(2)  & 0.92  & $3.3\times10^{-2}(2.2\times10^{-4})$ & (1)\\
4.6-5.6 & 42.5  & 1(1)  & 2.0   & $5.1\times10^{-4}(5.1\times10^{-4})$ & (1)\\
5-6     & 42.28 & 1  & 0.018 & $5.6\times10^{-3}$ & (2)\\
5.7     & 42.61 & 1  & 0.55  & $1.8\times10^{-4}$ & (3)\\
6.5     & 42.95 & 0  & 6.1   &  0                 & (4)\\
6.5     & 42.35 & 1  & 0.012 & $8.5\times10^{-3}$ & (4)\\
4.9     & 42.46 & 87 & 9.2   & $9.5\times10^{-4}$ & (5)\\
5.7     & 42.81 & 13 & 28    & $4.6\times10^{-5}$ & (6)\\
6.5     & 42.89 & 16 & 20    & $8.0\times10^{-6}$ & (7)\\
5.7     & 42.95 & 6  & 13    & $4.6\times10^{-5}$ & (8)\\
5.7     & 42.72 & 0  & 0.11  &  0                 & (9)\\
5.7     & 42.40 & 0  & 0.073 &  0                 & (9)\\
\tableline
$\sim5.8$  & 43.35 & 6  & 18    & $3.3\times10^{-5}$ & (10)\\
$\sim6.3$  & 43.32 & 26 & 1.7   & $1.5\times10^{-3}$ & (11)\\
$\sim5$    & 42.96 & 10 & 1.4   & $7.1\times10^{-4}$ & (12)\\
$\sim5.5$  & 43.40 & 2  & 2.4   & $8.3\times10^{-5}$ & (13)\\
$\sim5.5$  & 43.28 & 16 & 7.0   & $2.3\times10^{-3}$ & (13)\\
$\sim5.3$  & 43.18 & 6  & 12    & $5.0\times10^{-5}$ & (14)\\

\tableline
\end{tabular}
\tablenotetext{a}{Survey limiting $L$ in units of $\lumcgs$}
\tablenotetext{b}{Number of sources detected in survey}
\tablenotetext{c}{Volume of survey in units of $10^4~\mpc^3$}
\tablenotetext{d}{Number density of sources, in units of $\mpc^{-3}$}
\tablerefs{(1) this paper; (2) \citealt{hu98}; (3) \citealt{hu99}; (4)
\citealt{hu02a}; (5) \citealt{ouc03}; (6) \citealt{rho01,rho03}; (7)
\citealt{kod03}; (8) \citealt{mai03}; (9) \citealt{mar03}; (10)
\citealt{sta03}; (11) \citealt{yan03}; (12) \citealt{iwa03}; (13)
\citealt{fon03}; (14) \citealt{leh03}}
\label{tab:surveys}
\end{table}

\end{document}